\documentclass[fleqn,usenatbib]{mnras}

\usepackage[T1]{fontenc}

\DeclareRobustCommand{\VAN}[3]{#2}
\let\VANthebibliography\thebibliography
\def\thebibliography{\DeclareRobustCommand{\VAN}[3]{##3}\VANthebibliography}

\usepackage{graphicx}	% Including figure files
\usepackage{amsmath}	% Advanced maths commands
\usepackage{amssymb}	% Extra maths symbols

\usepackage{newtxtext,newtxmath}

\title[\textit{N}-body simulations of fSIDM]{\textit{N}-body simulations of dark matter with frequent self-interactions}

\author[M. S. Fischer et al.]{Moritz S. Fischer,$^{1}$\thanks{E-mail: moritz.fischer@uni-hamburg.de (UHH)}
Marcus Br\"{u}ggen,$^{1}$
Kai Schmidt-Hoberg,$^{2}$
Klaus Dolag,$^{3,4}$
\newauthor{Felix Kahlhoefer,$^{5}$
Antonio Ragagnin,$^{6,7}$
Andrew Robertson$^{8}$}
\\
$^{1}$Hamburger Sternwarte, Universit\"at Hamburg, Gojenbergsweg 112, D-21029 Hamburg, Germany\\
$^{2}$Deutsches  Elektronen-Synchrotron  DESY,  Notkestra{\ss}e  85,  D-22607  Hamburg,  Germany\\
$^{3}$Universit\"ats-Sternwarte M\"unchen, Faculty of Physics, Ludwig-Maximilians-Universität, Scheinerstr. 1, D-81679 M\"unchen, Germany\\
$^{4}$Max-Planck-Institut f\"ur Astrophysik, Karl-Schwarzschild-Str. 1, D-85748 Garching, Germany\\
$^{5}$Institute for Theoretical Particle Physics and Cosmology (TTK), RWTH Aachen University, D-52056 Aachen, Germany\\
$^{6}$INAF -- Osservatorio Astronomico di Trieste, via G. B. Tiepolo 11, I-34143 Trieste, Italy\\
$^{7}$Institute for Fundamental Physics of the Universe (IFPU), Via Beirut 2, I-34014 Trieste, Italy\\
$^{8}$Institute for Computational Cosmology, Durham University, South Road, Durham DH1 3LE, UK
}

\date{Accepted XXX. Received YYY; in original form ZZZ}

\pubyear{2020}

\begin{document}
\label{firstpage}
\pagerange{\pageref{firstpage}--\pageref{lastpage}}
\maketitle

\begin{abstract}
Self-interacting dark matter (SIDM) models have the potential to solve the small-scale problems that arise in the cold dark matter paradigm. Simulations are a powerful tool for studying SIDM in the context of astrophysics, but it is numerically challenging to study differential cross-sections that favour small-angle scattering, as in light-mediator models.
Here, we present a novel approach to model frequent scattering based on an effective drag force, which we have implemented into the \textit{N}-body code \textsc{gadget-3}.
In a range of test problems, we demonstrate that our implementation accurately models frequent scattering. Our implementation can be used to study differences between SIDM models that predict rare and frequent scattering. We simulate core formation in isolated dark matter haloes, as well as major mergers of galaxy clusters and find that SIDM models with rare and frequent interactions make different predictions. In particular, frequent interactions are able to produce larger offsets between the distribution of galaxies and dark matter in equal-mass mergers.
\end{abstract}


%%%%%%%%%%%%%%%%%%%%%%%%%%%%%%%%%%%%%%%%%%%%%%%%%%

%%%%%%%%%%%%%%%%% BODY OF PAPER %%%%%%%%%%%%%%%%%%

\section{Introduction}

Dark matter (DM) is an essential component of the standard cosmological model (Lambda cold dark matter, $\mathrm{\Lambda}$CDM), which was introduced to explain a variety of observations, such as the formation of large-scale structure and the cosmic microwave background.
These observations can be explained remarkably well under the assumption that DM is cold and collisionless \citep[e.g.][]{Planck_Collaboration_2020}. Nevertheless, on small scales, i.e.\ galactic scales, the predictions of $\mathrm{\Lambda}$CDM are in tension with observations.
The different aspects in which the predictions deviate from observations on small spatial scales could present a challenge to our $\mathrm{\Lambda}$CDM model.
Usually, up to five small-scale problems are considered. These are the missing satellites problem, the too-big-to-fail problem, the diversity problem, the core-cusp problem, and the plane-of-satellites problem \citep[for a review see][]{Bullock_2017}.
Not all of them describe actual problems of $\mathrm{\Lambda}$CDM and at least the missing satellites can be explained within the cosmological standard model \citep[e.g.][]{Kim_2018}.

In order to resolve the small-scale problems, a number of potential solutions have been proposed. Some of them attempt to mitigate the tensions by more accurate subgrid models of the baryonic physics in cosmological simulations.
It has been shown by numerous studies that DM cores can be created by feedback processes such as outflows from supernovae \citep{Read_2005, Governato_2012,Pontzen_2012,di_Cintio_2013, Brooks_2014, di_Cintio_2014, Pontzen_2014, Onorbe_2015, Tollet_2016, Benitez-Llambay_2019} as well as due to black holes \citep[e.g.][]{Martizzi_2013, Peirani_2017, Silk_2017}.
Other work shows that tensions can be reduced by improving the modelling of the internal dynamics of observed galaxies \citep{Oman_2019}.

An alternative to these small-scale problems pointing towards some deficiency in the modelling of baryons is that they are telling us something fundamental about the nature of DM. DM that is `warm' \citep{Dodelson_1994} or `fuzzy' \citep{Hu_2000} would change the abundance and internal structure of DM haloes.
Along this line, a promising alternative DM model is self-interacting dark matter (SIDM) \citep[for a review see][]{Tulin_2018}, which was proposed by \cite{Spergel_2000} as a solution to some of the small-scale problems.\footnote{Cold dark matter with self-interactions was first proposed by \cite{Carlsson_1992}, but this was in a different context.}

More precisely, SIDM is a class of many physics models that all have in common that DM consists of particles and these particles interact with each other so strongly that the interaction can alter the distribution of DM on astrophysical scales significantly, e.g.\ create density cores in haloes of DM.
From observations, the interesting range of cross-sections divided by DM mass is of the order of 1 cm$^2$ g$^{-1}$. In the limit of a negligible cross-section, SIDM behaves the same way as CDM.

A range of methods have been proposed to study the effects of SIDM on cosmic structures. The isothermal Jeans approach \citep{Kaplinghat_2014,Kaplinghat_2016} and the gravothermal fluid model \citep{Gnedin_2001, Balberg_2002a, Balberg_2002b, Koda_2011, Pollack_2015} are based on assuming that self-interactions maintain an equilibrium state, in which the full phase-space information is not required.
However, in many situations DM is neither collisionless nor fully collisional, which means that the simplifying assumption of local equilibrium cannot be made.
The most general -- but also computationally most expensive -- method to study SIDM is to run \textit{N}-body simulations. Here, the Vlasov--Poisson equation with a collision term for DM self-interactions is solved in six-dimensional phase-space.
The scattering of the numerical particles is treated analogously to physical particles.
The first such simulation using a Monte Carlo scheme for the scattering angle was performed by \cite{Burkert_2000}.
Modern schemes used for SIDM only differ from this approach in the way in which scattering probabilities are computed. 

The common approach of explicitly simulating individual scattering events leads to 
complications when the differential cross-section favours scattering by small angles. In this case it becomes necessary to simulate large numbers of scattering events that individually have negligible impact on the phase-space distribution. For very frequent scattering events, this becomes prohibitively expensive \citep{Robertson_2017b} because the required time-steps become too small.

Previous attempts to address this problem have relied on a number of simplifying assumptions. \cite{Kahlhoefer_2014} performed simplistic \textit{N}-body simulations of mergers by using an external gravitational potential for each halo and sampling the DM and galaxies with test particles. In this set-up, the effects of frequent self-interactions were modelled with an effective drag force.
\cite{Kummer_2019} combined the heat conduction approach from fluid models with \textit{N}-body simulations, which assumes that the system is in local equilibrium such that a well-defined temperature exists. Consequently, this approach is limited in its versatility, and, for instance, cannot be applied to merging systems. 

In this paper, we present a novel method designed for the frequent scattering regime, which enables general astrophysical simulations of frequent self-interacting dark matter (fSIDM). Our method employs a fundamentally different formulation of the collision term compared to the state-of-the-art schemes for rare self-interacting dark matter (rSIDM) and makes use of the fact that the effect of frequent scattering events in fSIDM can be described by an effective drag force \citep{Kahlhoefer_2014}.

In fSIDM, a DM particle travelling through a space filled with other DM particles would undergo many small-angle scattering events.
Each scattering event leads to a small change of the velocity, but the cumulative velocity change perpendicular to the initial direction of motion will tend to average out, with an expectation value: $\langle \delta v_\perp \rangle = 0$. The expected parallel component of the velocity change is non-zero ($\langle \delta v_\parallel \rangle \neq 0$), which can be interpreted as a drag force.
At the same time one finds $\langle \delta v_\perp^2 \rangle > 0$, which can be regarded as kinetic heating. 

Rare self-interactions have a much larger velocity change per scattering event and thus cannot, in general, be described by a drag force.
Only when the density is very high, an effective drag  can occur \citep{Kim_2017b}. As we will explore in this paper, the different effective descriptions of fSIDM and rSIDM lead to significant differences in the predicted DM distributions in astrophysical systems. In particular systems far away from equilibrium, such as ongoing major mergers of galaxy clusters respond in different ways to frequent and rare self-interactions. Indeed, we find that the effects for fSIDM can be substantially larger than those previously found for rSIDM.

This paper has several aims: First, we present a new numerical scheme to simulate frequent self-interactions of DM, which is described in Section~\ref{sec:numerical_method}. Secondly, in Section~\ref{sec:test_problems} we validate our scheme and its implementation in the $N$-body code \textsc{gadget-3} using a number of test problems. 
We then study core formation in an isolated Hernquist halo (see Section~\ref{sec:core-size}) and compare the effects of fSIDM and rSIDM. Finally, we explore differences between fSIDM and rSIDM in equal-mass mergers.
In Section~\ref{sec:merger_sim}, we simulate a merger of DM haloes with parameters typical of galaxy clusters.
In Section~\ref{sec:discussion}, we elaborate on various technical aspects of our code and discuss the physical implications of our results.
Finally, we summarize and conclude in Section~\ref{sec:conclusion}. Additional details are provided in the appendices.

\section{Numerical Method} \label{sec:numerical_method}

In this section, we first describe the key aspects of our method to simulate frequent scatterings. We then explain its implementation in the $N$-body code \textsc{gadget-3}, which contains a description of the parallelization. Finally, we point out differences between state-of-the-art schemes for rSIDM and our formulation of fSIDM.

In $N$-body codes, the phase-space distribution of DM is represented by numerical particles.
These particles each represent phase-space patches consisting of many physical particles. They are assigned a velocity and to smooth the represented matter distribution in configuration space a kernel is employed.
Such a Lagrangian description has some advantages over an Eulerian approach, such as Galilean invariance. But there are also disadvantages, for instance when it comes to the parallelization (see below).

Let us first look at the interaction of two phase-space patches, i.e.\ two numerical particles, which we assume to have equal mass. If the scattering is elastic, we can derive the post-scattering velocities of the particles from energy and momentum conservation: $E_i' + E_j' = E_i + E_j$ and $\mathbfit{p}'_i + \mathbfit{p}'_j = \mathbfit{p}_i +\mathbfit{p}_j$.

We divide the scattering process into two steps: The first one applies a drag force and the second one re-adds the energy lost in the first step. The latter is done in a random direction but perpendicular to the direction of motion to model kinetic heating. We indicate the intermediate state between the two steps by the superscript $^*$.
The velocity of the two numerical particles can be expressed as follows:
\begin{align} \label{eq:nf_velocity_1}
    \mathbfit{v}_i^{\ast} & = \mathbfit{v}_i - \Delta \mathbfit{v}_\mathrm{drag} \, , \qquad \mathbfit{v}_j^{\ast} = \mathbfit{v}_j + \Delta \mathbfit{v}_\mathrm{drag} \,,
\\
    \label{eq:nf_velocity_2}
    \mathbfit{v}_i{'} & = \mathbfit{v}_i^{\ast} + \Delta \mathbfit{v}_\mathrm{rand} \, , \qquad \mathbfit{v}_j{'} = \mathbfit{v}_j^{\ast} - \Delta \mathbfit{v}_\mathrm{rand} \,.
\end{align}
Here, $\Delta \mathbfit{v}_\mathrm{drag}$ denotes the velocity change due to the effective drag force. $\Delta \mathbfit{v}_\mathrm{rand}$ denotes the velocity which is added to ensure energy conservation, while momentum conservation is guaranteed as velocity changes are symmetric for the two particles.

\subsection{First step: apply drag force} \label{sec:apply_drag_force}

We introduce the relative velocity $\Delta \mathbfit{v}_{ij} = \mathbfit{v}_i - \mathbfit{v}_j$ and use it to express the velocity change $\Delta \mathbfit{v}_\mathrm{drag}$ due to the drag force,
\begin{equation}
    \Delta \mathbfit{v}_\mathrm{drag} = |\Delta \mathbfit{v}_\mathrm{drag}| \cdot \frac{\Delta \mathbfit{v}_{ij}}{|\Delta \mathbfit{v}_{ij}|} \,.
\end{equation}
Next, $|\Delta \mathbfit{v}_\mathrm{drag}|$ can be written as:
\begin{equation} \label{eq:vdrag_time_int}
    |\Delta \mathbfit{v}_\mathrm{drag}| = \frac{F_\mathrm{drag}}{m} \cdot \Delta t \,.
\end{equation}
The drag force is given by $F_\mathrm{drag}$ and $\Delta t$ denotes the time-step. 

To derive the drag force, we start from the same assumptions as made by \cite{Kahlhoefer_2014}. They derived the deceleration rate ($R_\mathrm{dec} \equiv v_0^{-1} \, \mathrm{d}v_\parallel / \mathrm{d}t$) of an individual physical DM particle travelling with velocity $v_0$ through a background density $\rho_j$ and found
\begin{align} \label{eq:decel_rate}
    R_\mathrm{dec} = \frac{\rho_j \, v_0 \, \sigma_\mathrm{\tilde{T}}}{2\,m_\chi}\,.
\end{align}
Here, $m_\chi$ denotes the mass of a DM particle and 
\begin{equation}
    \sigma_\mathrm{\tilde{T}} = 4\pi \int_0^1 \frac{\mathrm{d} \sigma}{\mathrm{d} \Omega_\text{cms}} (1 - \cos \theta_\text{cms}) \mathrm{d} \cos \theta_\text{cms} 
\end{equation}
denotes the momentum transfer cross-section.\footnote{Note that if the differential cross section is invariant under the exchange $\theta \to \pi - \theta$ (as in the case of the scattering of identical particles), this definition is equivalent to the one advocated by \cite{Robertson_2017b,Kahlhoefer:2017umn}: $\sigma_\mathrm{\tilde{T}} = 2\pi \int_{-1}^1 \frac{\mathrm{d} \sigma}{\mathrm{d} \Omega_\text{cms}} (1 - |\cos \theta_\text{cms}|) \mathrm{d} \cos \theta_\text{cms}$.
}
In the regime of isotropic scattering, this definition is a factor of 2 smaller than the one commonly used in studies of rSIDM.

To apply this calculation to our simulations we interpret the background density $\rho_j$ as the density of a single phase-space patch represented by a numerical particle. Moreover, we need to consider the scattering of many particles and their total momentum change, which can be written as
\begin{equation}
    \mathrm{d} p_\parallel = \int n_i \, m_\chi \, \mathrm{d}v_\parallel \mathrm{d}V = \int \rho_i \, \mathrm{d}v_\parallel \mathrm{d}V \,.
\end{equation}
Here, the number density of physical DM particles belonging to phase-space patch $i$ is given by $n_i = \rho_i / m_\chi$, where $\rho_i$ denotes the mass of the numerical particle multiplied by the kernel: $\rho_i(\mathbfit{x}) = m_i \cdot W(|\mathbfit{x}-\mathbfit{x}_i|,h_i)$ with $h_i$ being the kernel size (see below). The physical density can then be obtained by summing over all numerical particles at a given position.

Using the deceleration rate from equation~\eqref{eq:decel_rate} we can therefore express the resulting drag force acting on a phase-space patch as
\begin{equation} \label{eq:drag_force}
    F_\mathrm{drag} = \frac{1}{2} \, |\Delta\mathbfit{v}_{ij}|^2 \, \frac{\sigma_\mathrm{\tilde{T}}}{m_\chi} \int \rho_i \, \rho_j \, \mathrm{d}V \,.
\end{equation}
Using the kernel $W(|\mathbfit{x}-\mathbfit{x}_i|,h_i)$, we can express the drag force as
\begin{align} \label{eq:drag_force2}
    F_{\mathrm{drag}} =& \, \frac{1}{2} \, |\Delta\mathbfit{v}_{ij}|^2 \, \frac{\sigma_\mathrm{\tilde{T}}}{m_\chi} \, m_i \, m_j \nonumber \\
    &\cdot \int W(|\mathbfit{x}-\mathbfit{x}_i|,h_i) \cdot W(|\mathbfit{x}-\mathbfit{x}_j |,h_j) \, \mathrm{d}\mathbfit{x} \,.
\end{align}
Note that we do not consider interactions between particles belonging to the same phase-space patch, as they have parallel trajectories. Furthermore, we assume $\sigma_\mathrm{\tilde{T}}$ to be velocity independent in this work.

\subsection{Second step: re-add energy}

In the second step, we re-add the energy $\Delta E$ that is lost due to the drag force.
It can be written as
\begin{equation} \label{eq:energy_drag}
    \frac{2 \, \Delta E}{m} = |\Delta \mathbfit{v}_\mathrm{drag}| \, \left( |\Delta \mathbfit{v}_{ij}| - |\Delta \mathbfit{v}_\mathrm{drag}|\right) \,.
\end{equation}
To ensure that the local velocity distribution evolves towards thermal equilibrium, the added velocity $\Delta \mathbfit{v}_\mathrm{rand}$ needs to be perpendicular to the relative velocity $\Delta \mathbfit{v}^{\ast}$.
We then find that
\begin{equation} \label{eq:velocity_rand}
    |\Delta \mathbfit{v}_\mathrm{rand}| = \sqrt{\frac{2 \Delta E}{m}} \,.
\end{equation}
The direction of $\Delta \mathbfit{v}_\text{rand}$ is chosen randomly in the plane orthogonal to $\Delta \mathbfit{v}_{ij}$.
Once the velocity change due to the random component has been computed, we can update the velocity according to equation~\eqref{eq:nf_velocity_2} and continue with the next particle pair.

Note, that the post-scattered velocities $\mathbfit{v}'$ are treated as pre-scattered velocities $\mathbfit{v}$ for any subsequent pair computations in the same time-step. This implies that the result depends on the exact order in which the particle pairs are considered. However, this is an effect which is only relevant at the level of individual particle trajectories. It has no meaning for the statistical properties of the DM distribution. Treating particle pairs in a different order would lead to a different \textit{N}-body representation of the same distribution, as would different random directions for the re-added energy. 

\subsection{Kernel}

The drag force computation is based on a kernel function representing the DM density distribution of a numerical particle in configuration space. Here, we will discuss the use of kernel functions, describe how we compute the overlap and explain how we choose the kernel size.

The application of kernel functions in this work is quite different from the one in smoothed particle hydrodynamics (SPH), where they are used to compute derivatives and therefore need to be differentiable. For the scheme presented here, we only need to integrate the kernel functions as described in Section~\ref{sec:apply_drag_force}.
We have tried a number of different kernel functions and found that they all perform similarly well in the context of the first test problem presented in sec.~\ref{sec:test_problem_deceleration}.
In the end, we choose the spline kernel introduced by \cite{Monaghan_1985}, which is very popular in SPH.
For our studies, we use a scaled version such that it becomes zero for $r \geq h$, where $h$ denotes the kernel size:
\begin{equation}
    W(r,h)= \left\lbrace \begin{matrix} \frac{8}{\pi \, h^3} \left[1-6 \, (r/h)^2 \, (1- r/h)\right]&\quad \textnormal{if} &0\leq r/h < 0.5,\\ \frac{16}{\pi h^3}[1-(r/h)]^3&\quad \textnormal{if}&0.5\leq r/h < 1, \\ 0 &\quad \textnormal{if}& 1 \leq r/h . \end{matrix} \right.
\end{equation}

Using this kernel, we can calculate the overlap $\Lambda_{ij}$ of the particles $i$ and $j$, which corresponds to the integral of equation~\eqref{eq:drag_force2}:
\begin{equation} \label{eq:overlap}
    \Lambda_{ij} = \int W(|\mathbfit{x}-\mathbfit{x}_i|,h_i) \cdot W(|\mathbfit{x}-\mathbfit{x}_j |,h_j) \, \mathrm{d}\mathbfit{x} \, .
\end{equation}
Details on how this integral is calculated in practice, are given in Appendix~\ref{app:kernel_overlap}. 

The kernel size should be chosen adaptively to reach a high resolution in regions with a large particle number density. Simultaneously, the kernel size needs to be large in low-density regions to ensure that each particle has a sufficient number of neighbours to interact with. We use the common method to set the kernel size to a length such that the kernel includes a given number of neighbouring particles $N_\mathrm{ngb}$. 

\subsection{Time-step}

Our implementation of frequent scattering does not introduce an additional time-step constraint.
This is because for our simulations the gravitational time-step is smaller than what is required for the frequent self-interactions. However, for different applications it is conceivable that the gravitational time-step becomes larger than what is needed for the scattering, for example in the phase of gravothermal collapse of a DM halo. Let us therefore briefly outline how the time-step requirements depend on the relevant quantities.

For the derivation of a time-step criterion one can start from the drag force:
\begin{equation}
    F_\mathrm{drag} = \frac{1}{2} \, |\Delta v|^2 \, \frac{\sigma_\mathrm{\tilde{T}}}{m_\chi} \, m^2 \, \Lambda \, ,
\end{equation}
where $m$ denotes the simulation particle mass and we have dropped the subscripts $i,j$.
This equation implies a velocity change of
\begin{equation} \label{eq:timestep1}
    \Delta v_\mathrm{drag} = \frac{F_\mathrm{drag}}{m} \Delta t = \frac{1}{2} \, |\Delta v|^2 \, \frac{\sigma_\mathrm{\tilde{T}}}{m_\chi} \, m \, \Lambda \, \Delta t \, .
\end{equation}
Here, $\Delta t$ denotes the time-step.
For a conservative estimate we replace $m \, \Lambda$ with $\tilde{\rho}$ , which is inversely proportional to $N_\mathrm{ngb}$:
\begin{equation} \label{eq:timestep2}
    \tilde{\rho} = \frac{3 \, m}{4 \pi \, h^3} \sim \frac{\rho}{N_\mathrm{ngb}}\,.
\end{equation}
From equations~\eqref{eq:timestep1} and \eqref{eq:timestep2}, we derive the time-step assuming the numerical error is kept constant ($\Delta v_\mathrm{drag} / |\Delta v| = \mathrm{const}$):
\begin{equation} \label{eq:time_step_scaling}
    \frac{1}{\Delta t} \sim |\Delta v| \, \frac{\sigma_\mathrm{\tilde{T}}}{m_\chi} \, \frac{\rho}{N_\mathrm{ngb}} \, .
\end{equation}

One finds that a smaller time-step is required when using a larger SIDM cross-section or when the relative velocities, i.e. the velocity dispersion, increases. Moreover, a smaller time-step is reasonable in dense regions.
From equation~\eqref{eq:time_step_scaling}, we also obtain a dependence on the number of neighbours, choosing a larger value can relax the time-step constraint.

\subsection{Implementation in \textsc{gadget-3}}

We implement the DM self-interactions in \textsc{gadget-3}, which is an updated version of the \textit{N}-body code \textsc{gadget-2} \citep{gadget2}.\footnote{Recently, the latest version \textsc{gadget-4} has been published \citep{gadget4}.} Here, we will describe our implementation in the simulation code. We begin by describing how to find pairs of particles that should interact. Then, we comment on adaptive time-stepping. Lastly, we explain how we deal with the largest challenge posed by the parallelization. 

\subsubsection{How to find interacting particles?} \label{sec:how_to_find_interacting_particles}
In \textsc{gadget-3} a tree structure is used in the gravity calculation, and we use this same tree to find neighbouring particles that will scatter with one another. Defining the distance between particles $i$ and $j$ as $d_{ij}$, we use the tree to find all particle pairs for which $d_{ij} < h_i + h_j$.
For all pairs of particles that fulfil this relation, we compute the effect of the frequent self-interactions and apply the velocity changes. Note, that for particles separated by more than the sum of their kernel sizes the overlap expressed by equation~\ref{eq:overlap} vanishes.

\subsubsection{Adaptive time-stepping} \label{sec:adaptive_time_step}
\textsc{gadget-3} uses an adaptive time-stepping scheme, where individual time-steps are assigned to each particles, with a power-of-two hierarchy of time-step lengths. Our scheme for frequent self-interactions is not based on individual particles, but on pairs of particles. Consequently, we need to compute a time-step for a pair.

The adaptive time-stepping scheme assigns particles to time-step bins, which leads to active and passive particles. The details can be found in the \textsc{gadget-2} paper \citep{gadget2}. In consequence, a pair consists of one active particle and one which is active or passive. For an active--active pair, the time-step of the pair is given by
\begin{equation}
    \Delta t = \frac{\min(\Delta t_i, \Delta t_j)}{2} \, ,
\end{equation}
where we divide by two because active--active pairs are considered twice per time-step (i.e.\ particle $i$ finds particle $j$ as a neighbour and vice versa). In the active--passive case, the pair is considered only once per time-step. Assuming that the active particle has the index $i$, the time-step can be written as
\begin{equation}
    \Delta t = \Delta t_i \, .
\end{equation}
We wish to point out that the time-step of the active particle is always shorter than that of the passive one.
The time-step $\Delta t$ computed as described above is used to compute the change in velocity due to the drag force using equation~\eqref{eq:vdrag_time_int}.

\subsubsection{Parallelization} \label{sec:parallelisation}

The parallelization of our scheme for frequent self-interactions is more complicated than for classical gravity or hydrodynamic schemes. The difficulties arise from the fact that one cannot treat the velocity change due to the particle--particle self-interactions cumulatively. Rather the computation of a scattering event requires the information from previous scatterings. Consequently, we cannot send one particle to multiple processes (execution instances of a computer program) simultaneously to make sure that each particle is only used by one process at a time. In addition, when sending particles to other processes it needs to be ensured that they are not needed locally (by the sending process) to scatter with particles received from other processes.  This is ensured by allowing only half of the processes to send particles at a time, while the other half only receives particles. Consequently, only the processes that receive particles compute the scattering, while the other half of the processes wait.

The communication between the processes is done in multiple sub-steps. We allow every process to communicate with all the other processes, but only one per sub-step. Given $N$ processes we have $B=N-1$ sub-steps.
In each of these sub-steps, we create pairs of processes and the two processes of a pair communicate with each other, i.e.\ exchange particle data. The pairs of a sub-step do not have common members, i.e.\ they are disjoint sets.
In practice, we have $2B$ sub-steps, i.e.\ every pair is considered twice. The first $B$ sub-steps are used for sending particles to the process of a pair that has the larger ID (a unique number for identification) and in the second $B$ sub-steps data is sent to the process with the smaller ID. Theoretically, sending particles to only one process of a pair could be enough, i.e.\ having $B$ sub-steps. But in practice, it is more complicated than the exchange in both directions due to the use of adaptive time-stepping. The local process $p$ given a sub-step $b$ communicates with $c = p \oplus b$. Here, $b \in [1,B]$ and $\oplus$ denotes the XOR operator. This scheme has the advantage that it can be easily implemented. However, it does not give the best performance theoretically possible because half of the processes are waiting while the non-local scattering is computed and also because symmetries are not exploited, i.e.\ each process pair is considered twice per time-step. Nevertheless, the parallelization leads to a large speed-up of the computations and thus allows us to run reasonably large simulations.

This parallelization scheme can also be used for infrequent large-angle scattering.
It allows overcoming the problem of `bad scatterings' observed by \cite{Robertson_2017a}, although it is more computationally expensive because each process can only communicate with one other process at a time, requiring more communication cycles per simulation time-step.
Our implementation of rare scattering is described in Appendix~\ref{sec:rSIDM_implementation}.

\subsection{Differences to numerical modelling of infrequent scattering}

To conclude the presentation of our numerical method let us discuss the differences to the common Monte-Carlo scheme for large-angle scattering.
The modelling of such infrequent scattering events with the \textit{N}-body method has similarities to our approach described above in the sense that both methods are based on the same numerical representation, but they are not identical. In the following, we point out differences referring to the scheme used by \cite{Rocha_2013}.

First of all, the scheme for infrequent scattering computes a probability that two particles with a separation smaller than the kernel size scatter. This is in contrast to the presented scheme for frequent interactions, where a drag force acts on all particle pairs with a sufficiently small separation, i.e.\ overlapping kernel functions.

Furthermore, the two schemes differ in the magnitude and the direction of the velocity change. For the infrequent scattering the post-scattering velocity of particle $1$ interacting with particle $2$ can be expressed as
    \begin{equation}
        \mathbfit{v}_1{'} = \mathbfit{v}_1 - \Delta \mathbfit{v} + |\Delta \mathbfit{v} | \cdot \mathbfit{e}_r \qquad \textnormal{with} \qquad \Delta \mathbfit{v} = \frac{\mathbfit{v}_1 - \mathbfit{v}_2}{2} \,,
    \end{equation}
    where $\mathbfit{e}_r$ is a random direction. 
    The corresponding equation for the frequent scattering scheme is given by
    \begin{equation}
        \mathbfit{v}_1{'} = \mathbfit{v}_1 - \Delta \mathbfit{v}_\mathrm{drag} + |\Delta \mathbfit{v}_\mathrm{rand}| \cdot \mathbfit{e}_f \,,
    \end{equation}
where $\mathbfit{e}_f$ denotes a random direction perpendicular to $\Delta \mathbfit{v}_\mathrm{drag}$. Crucially, $|\Delta \mathbfit{v}_\mathrm{drag}|, |\Delta \mathbfit{v}_\mathrm{rand}| \ll |\Delta \mathbfit{v}|$, i.e.\ the velocity of the scattering particles change only slightly in fSIDM, while the differences can be of order unity in rSIDM.

Besides, the rSIDM scheme provides a more general description of self-interactions and is also capable of describing highly anisotropic cross-sections, when $\mathbfit{e}_r$ is chosen according to the differential cross-section. But for those cross-sections favouring small-angle scattering, it would require a very large number of individual scattering events, which would cause a problem in terms of run time.

\section{Verification tests} \label{sec:test_problems}

To test that our numerical scheme works properly and that the implementation accurately models frequent self-interactions, we use several test-problems, which we present in this section.
The first problems study purely self-interactions. In contrast, the last problem, where we simulate an isolated DM halo, is motivated by astrophysics and includes gravity.

\subsection{Deceleration problems} \label{sec:test_problem_deceleration}

In our first test problem, we study a particle travelling through a background density, which is sampled by particles at rest.
Here, we only consider the drag force and neglect the random component.
Due to the drag force the test particle, which has a non-zero initial velocity, is decelerated by the background particles.
We compare the trajectory of the test particle to the exact solution, obtained from
\begin{equation}
    \Ddot{x} =-\frac{1}{2} \, \Dot{x}^2 \, \rho \, \frac{\sigma_\mathrm{\tilde{T}}}{m_\chi} \, .
\end{equation}

We make use of two different initial conditions. First, we consider a constant density and secondly, we introduce a density gradient.
For both we use $10^4$ particles. They have a total mass of $10^{10} \, \mathrm{M_\odot}$. A self-interaction cross-section of $\sigma_\mathrm{\tilde{T}}/m = 200 \, \mathrm{cm}^2 \, \mathrm{g}^{-1}$ is used for the test simulations and the time-step is set to $\Delta t = 0.02 \, \mathrm{Gyr}$.
Besides, $N_\mathrm{ngb} = 64$ is used to determine the size of the spline kernel, which is used to compute the drag force.

\subsubsection{Without density gradient}

\begin{figure}
    \centering
    \includegraphics[width=\columnwidth]{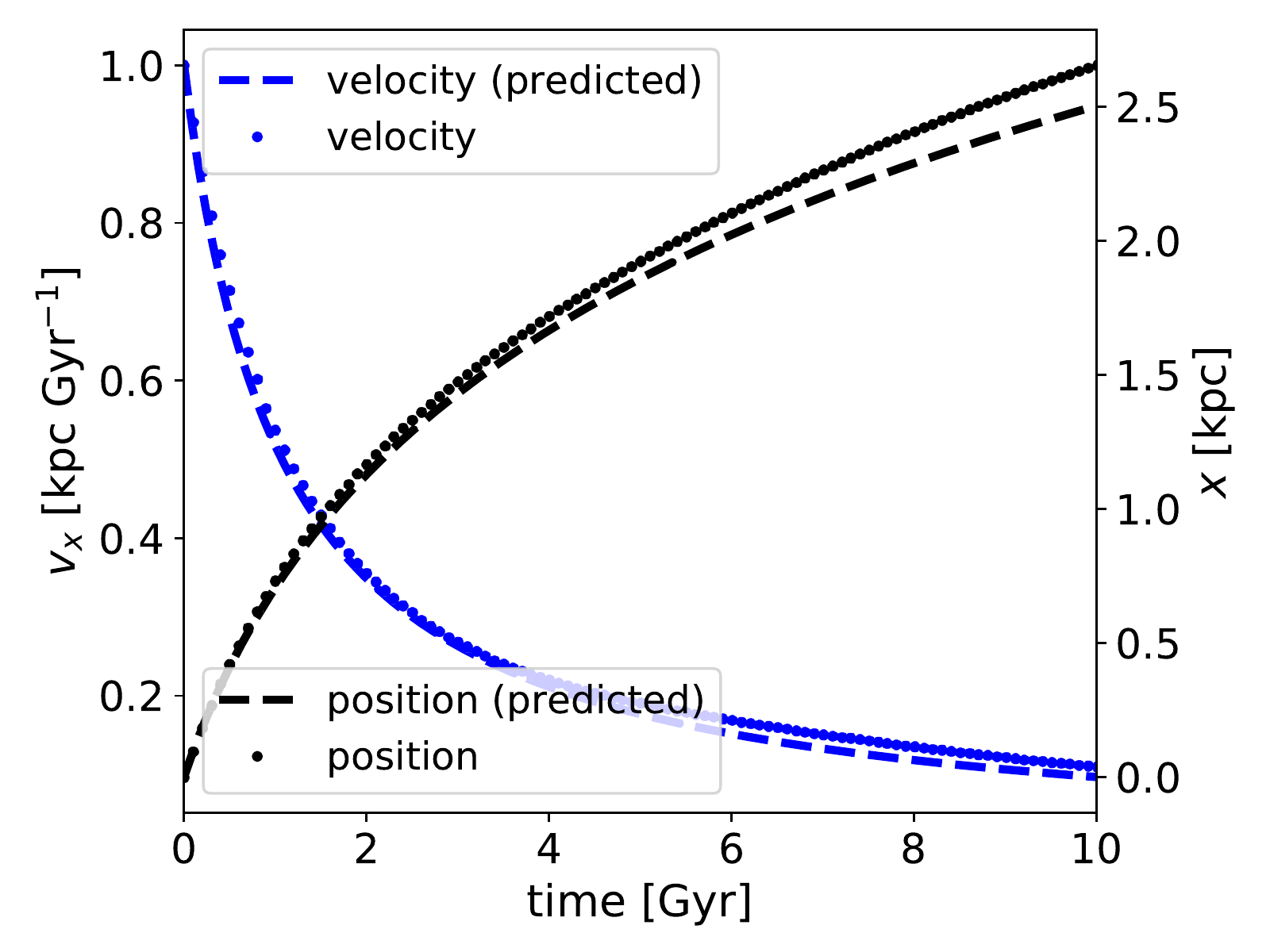}
    \includegraphics[width=\columnwidth]{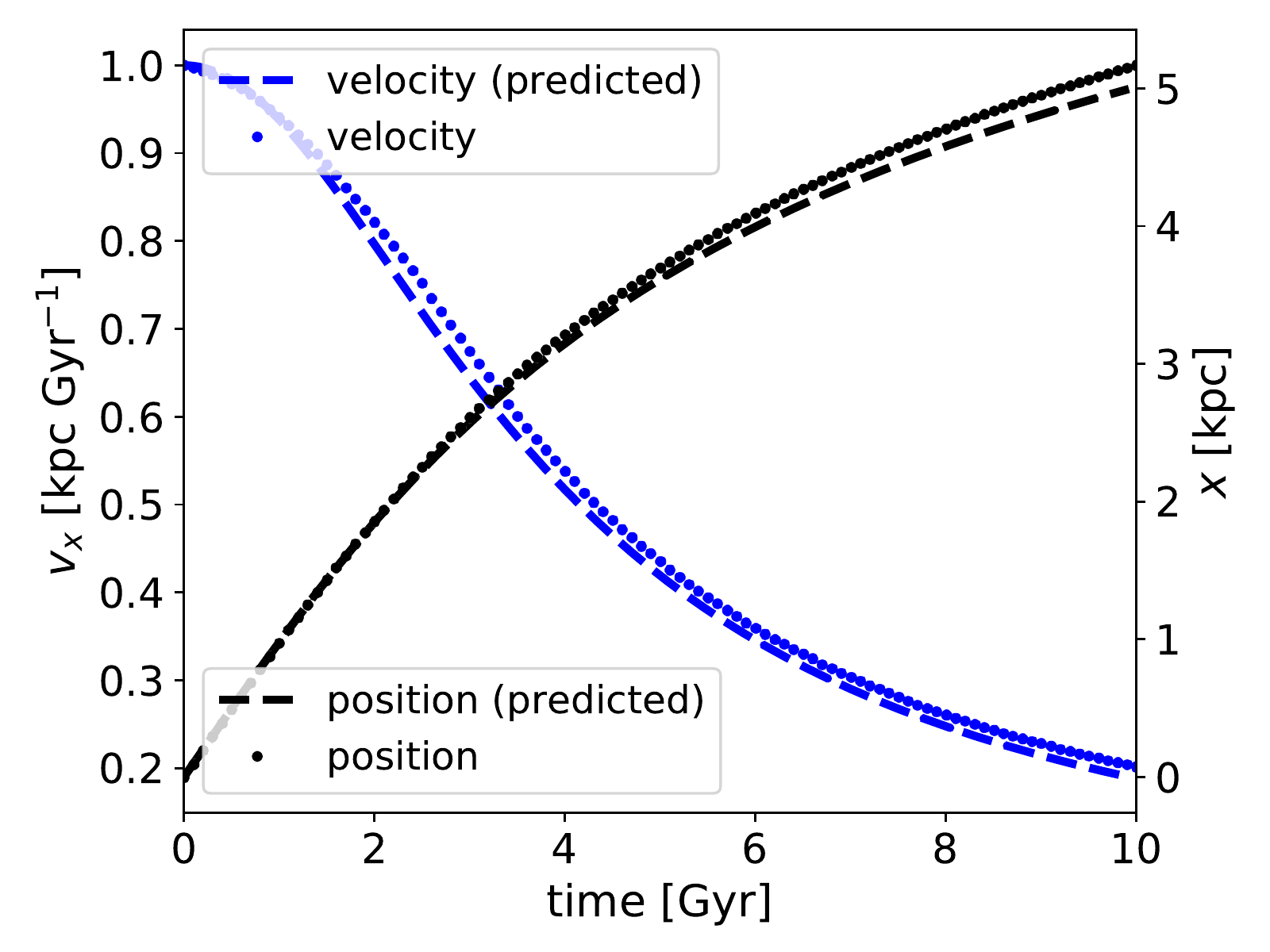}
    \caption{A particle is travelling through a constant background density (upper panel) or a linear background density with positive gradient (lower panel) and is decelerated through DM self-interactions. A velocity-independent cross-section of $\sigma_\mathrm{\tilde{T}}/m = 200 \, \mathrm{cm}^2 \, \mathrm{g}^{-1}$ is applied. The desired number of neighbours is set to 64.
    }
    \label{fig:gadget_test_vel_single}
\end{figure}

First, we choose a constant background density with an average density of $4.46 \times 10^{7} \, \mathrm{M_\odot} \, \mathrm{kpc}^{-3}$. In Fig.~\ref{fig:gadget_test_vel_single} (upper panel), we show the velocity (blue) and position (black) of the test particle.
For the velocity we find only minor deviations which should be negligible. For the particle position, the deviation is the integral of the minor deviations in the velocity.
Here, we find a larger deviation at the end of the simulations.
However, we do not expect this numerical error to be of a problematic size.
Later we comment on the accuracy and argue that we expect a higher accuracy for typical astrophysical simulations. 

\subsubsection{With density gradient}
Secondly, we choose a linear background density.
The density is zero at the initial location of the test particle and increases linearly along its path. 
The simulation results are shown in Fig.~\ref{fig:gadget_test_vel_single} (lower panel). The exact solution is computed numerically using a Runge--Kutta fourth-order method.
Compared to the problem without density gradient (Fig.~\ref{fig:gadget_test_vel_single}, upper panel) we find even smaller deviations from the exact solution.

It is worth mentioning that in a typical astrophysical simulation of fSIDM the relevant self-interaction cross-sections are smaller than the one simulated here by at least one order of magnitude, while the typical DM densities are comparable.
Moreover, in astrophysical simulations the time-steps will usually be much smaller because of the gravity constraints.
Both will increase the accuracy of the modelling of frequent interactions.

\subsection{Thermalization problem}
In this test problem, we study a periodic box that contains randomly distributed particles.
Initially, all particles have the same absolute velocity but with random orientation.
The system is not in equilibrium but is expected to evolve towards an equilibrium state.
The velocity distribution should become Maxwellian due to the self-interactions.

For the simulation we used $10^4$ particles representing a total mass of $10^{10} \, \mathrm{M_\odot}$ within a cubic box of a side length of $10\,\mathrm{kpc}$, the corresponding density is $10^{7} \, \mathrm{M_\odot} \, \mathrm{kpc}^{-3}$.
Initially, the absolute velocity of all particles is set to $2 \, \mathrm{kpc} \, \mathrm{Gyr}^{-1}$.
We use a self-interaction cross-section of $\sigma_\mathrm{\tilde{T}}/m = 10 \, \mathrm{cm}^2 \, \mathrm{g}^{-1}$, a time-step of $\Delta t = 0.012 \, \mathrm{Gyr}$ and $N_\mathrm{ngb} = 64$.

\begin{figure}
    \centering
    \includegraphics[width=\columnwidth]{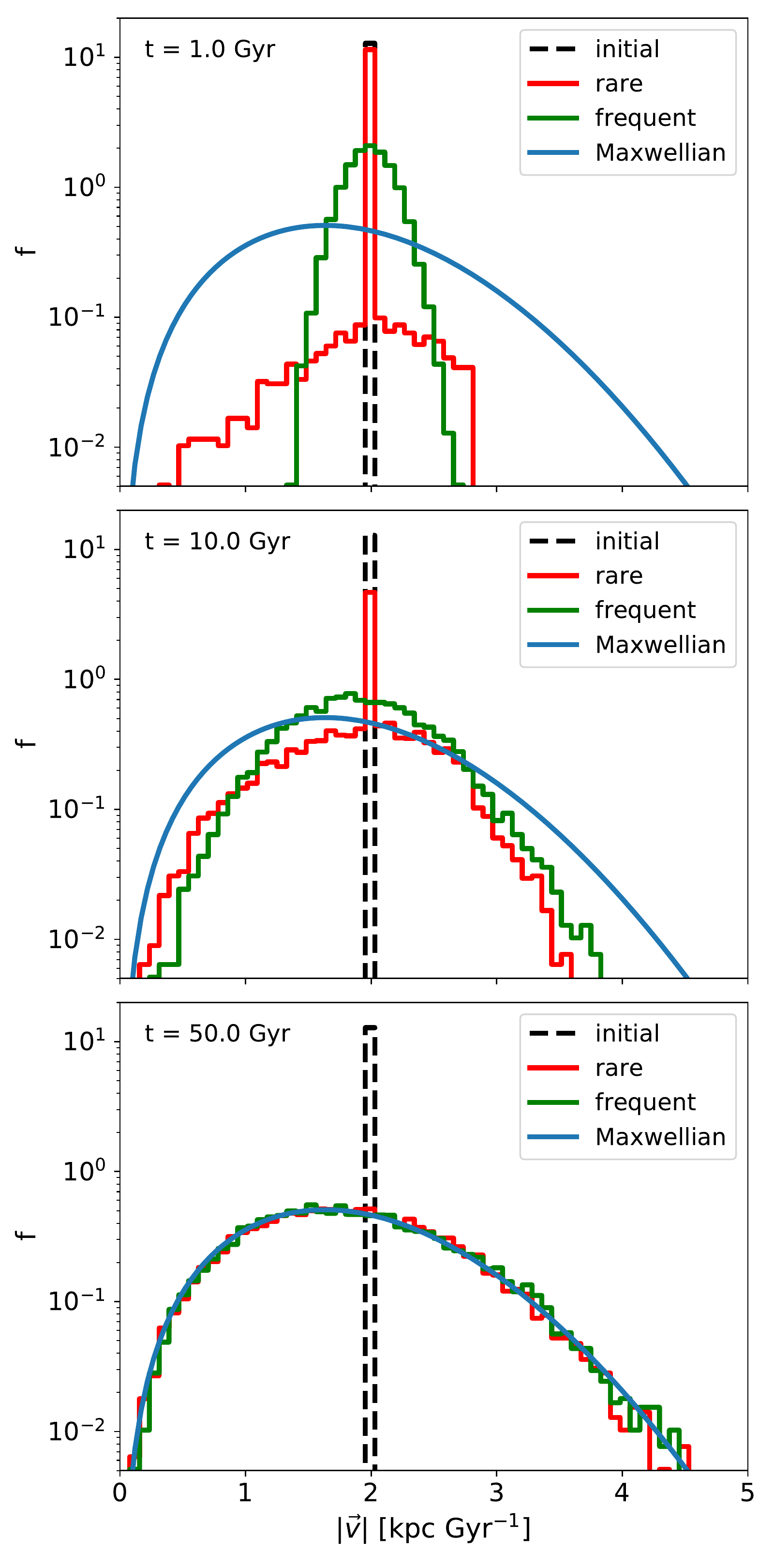}
    \caption{The velocity distributions for the thermalization problem are shown. The initial distribution is given in black. The evolution for rare (red) and frequent (green) self-interactions is shown for $t = 1$ Gyr (upper panel), $10$ Gyr (middle panel) and $50$ Gyr (lower panel). The plots demonstrate that the system evolves towards a Maxwell--Boltzmann distribution. The expected Maxwellian is plotted as well. In total 10000 particles were simulated with a cross-section of $\sigma_\mathrm{\tilde{T}}/m = 10 \, \mathrm{cm}^2 \, \mathrm{g}^{-1}$.}
    \label{fig:gadget_test_vel_dist2}
\end{figure}

In Fig.~\ref{fig:gadget_test_vel_dist2}, we show our results for this test problem.
We simulated the test problem with rare and frequent self-interactions.
Our implementation of rare scattering is described in Appendix~\ref{sec:rSIDM_implementation}.
Indeed, for both fSIDM and rSIDM we ultimately obtain a Maxwellian velocity distribution which is stable over time (lower panel).
However, the shapes of the intermediate velocity distributions (upper and middle panels) are quite different for the two cases.
The velocity distribution peak of rare self-interactions at $2\,\mathrm{kpc} \, \mathrm{Gyr}^{-1}$ is mainly due to unscattered particles. The sharp cut at large velocities after 1 Gyr (upper panel) can be explained by the maximum velocity that a particle can gain due to a single scattering event, $v_\mathrm{max} = \sqrt{2} \, v_\mathrm{ini}$.
The distribution function can become non-zero beyond that limit only if particles scatter multiple times.
The middle panel reveals that rare self-interactions lead to more particles in the low-velocity regime, whereas frequent interactions produce more high-velocity particles. 

\subsection{Angular deflection problem}
\label{sec:test_problem_deflection}
Our last test problem that purely studies the frequent self-interactions deals with a particle travelling through a constant background density.
Along its path, the particle undergoes many small-angle scattering events and gets deflected.
We measure the total deflection angle of many particles and compare them to the probability density function of Moli\`{e}re's theory \citep{Moliere_1948}\footnote{For a paper written in English on Moli\`{e}re's theory, we refer to \cite{Voskresenskaya_2012}.}.

We are simulating 8000 test particles with an additional 92$\,$000 particles to model the density background.
In total, the simulation contains a mass of $10^{10} \, \mathrm{M_\odot}$, which resides in a cube with a side length of $14 \, \mathrm{kpc}$.
This implies a background density of $\rho = 3.353 \times 10^{6} \, \mathrm{M_\odot} \, \mathrm{kpc}^{-3}$.
The initial velocity of the test particles is $v_\mathrm{init} = 2.0 \, \mathrm{kpc} \, \mathrm{Gyr}^{-1}$, while the background particles are at rest.
For the simulation we use a cross-section of $\sigma_\mathrm{\tilde{T}} = 10 \, \mathrm{cm}^2 \, \mathrm{g}^{-1}$, a time-step of $\Delta t = 0.001 \, \mathrm{Gyr}$ and $N_\mathrm{ngb}=64$.
The deflection angle $\theta$ of the test particles is defined as the angle between the initial and the current velocity vectors in the centre-of-mass system of the scattering physical particles, i.e.\ where they have the initial velocity of $v_\mathrm{init}/2$.
The details about the derivation of the prediction from Moli\`{e}re's theory can be found in Appendix~\ref{ap:moliere}.

In Fig.~\ref{fig:gadget_test_deflection_angle}, we show our results for the distribution of the deflection angles.
The left-hand panel shows the distribution after the particles have travelled $0.01 \, \mathrm{Gyr}$ within the target and the right one is for $t = 0.1 \, \mathrm{Gyr}$.
The plots demonstrate that our simulation agrees well with Moli\`{e}re's theory.
From the test problems studied so far, we can conclude that we are able to model frequent self-interactions accurately. 

\begin{figure}
    \centering
    \includegraphics[width=\columnwidth]{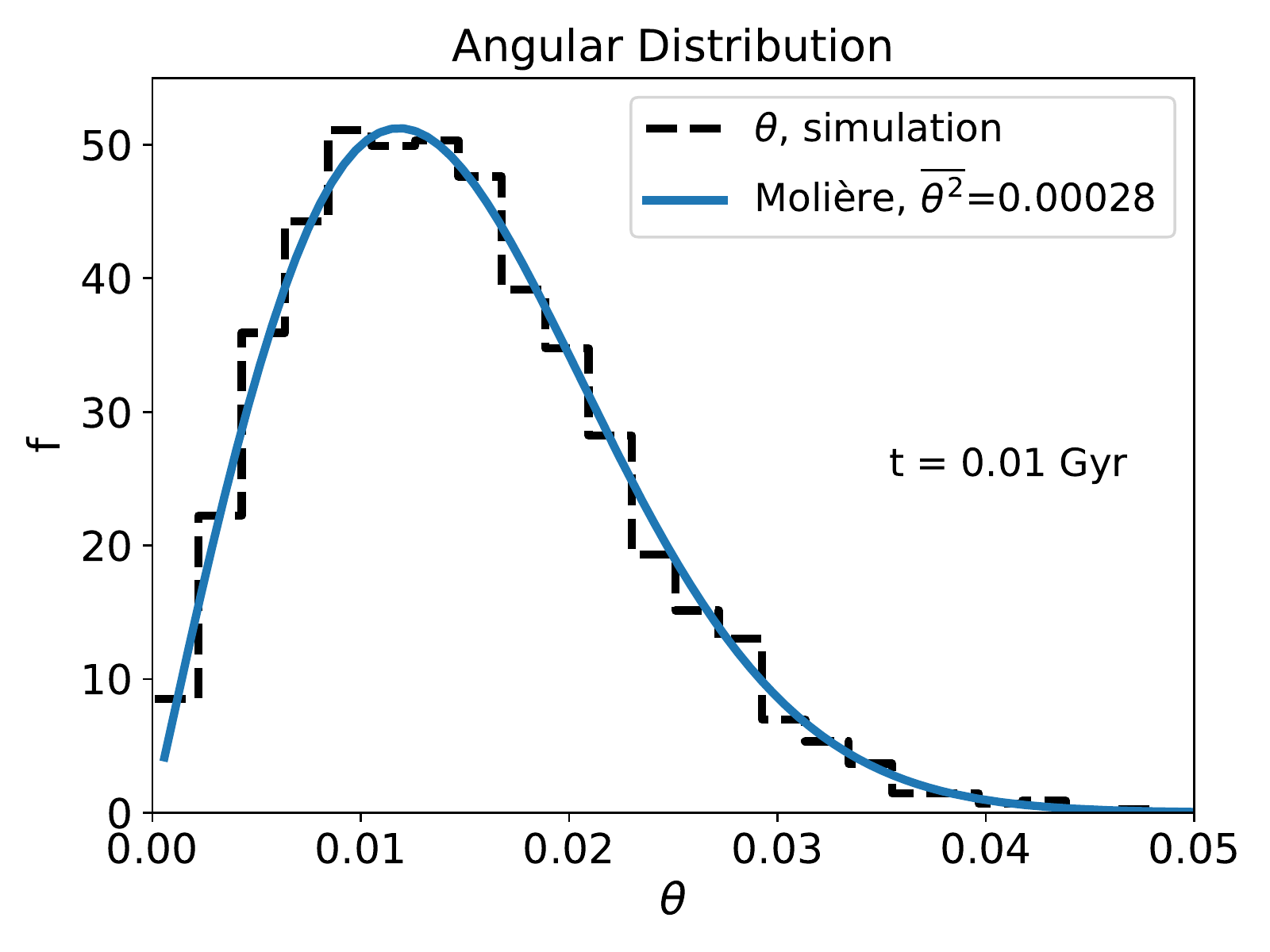}
    \includegraphics[width=\columnwidth]{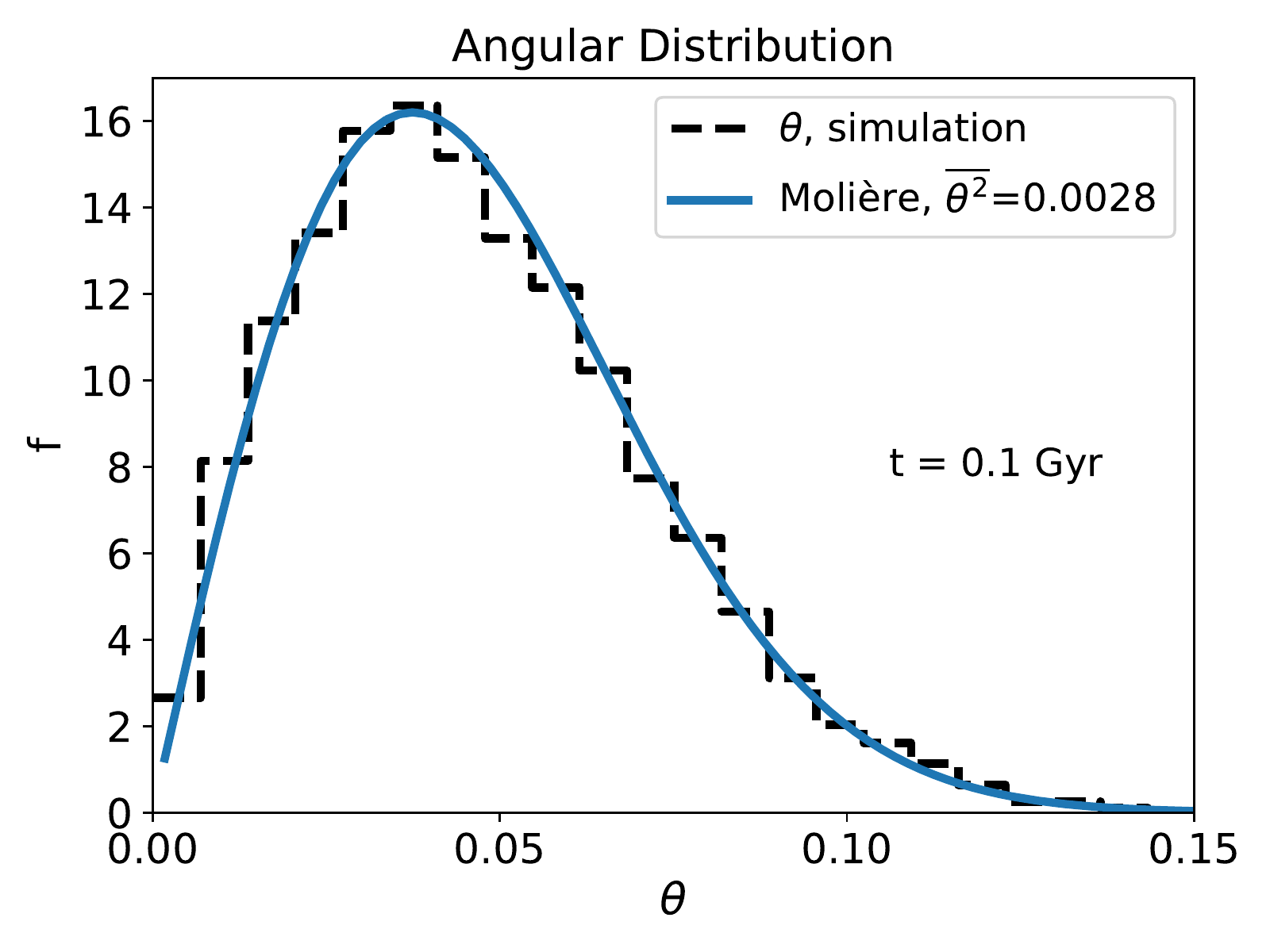}
    \caption{The distribution of the total deflection angle after 0.01 Gyr (upper panel) and 0.1 Gyr (lower panel). The self-interaction cross-section was chosen as $\sigma_\mathrm{\tilde{T}}/m = 10 \, \mathrm{cm}^2 \, \mathrm{g}^{-1}$. A number of 8000 test particles were used and in total 100$\,$000 particles were simulated.}
    \label{fig:gadget_test_deflection_angle}
\end{figure}

\subsection{NFW halo} \label{sec:halo_sim}

To test our scheme for frequent self-interactions in an astrophysical context including gravity we simulate an isolated DM halo.
As initial condition we choose a halo with a  Navarro--Frenk--White (NFW) profile \citep{Navarro_1996} with $M_\mathrm{vir} = 10^{15} \, \mathrm{M_\odot}$, $r_\mathrm{s} = 300 \, \mathrm{kpc}$, and $\rho_\mathrm{s} \equiv \rho(r_s) = 7.25 \times 10^5 \, \mathrm{M_\odot} \, \mathrm{kpc}^{-3}$.
The NFW halo is sampled up to the virial radius ($r_\mathrm{vir} = 1626 \, \mathrm{kpc}$).
We integrate the Jeans equation to obtain the velocity dispersion. To sample the initial velocities, we locally approximate the velocity distribution by a Maxwell--Boltzmann distribution, i.e.\ we draw the velocity components randomly from a Gaussian.
The gravitational softening length is set to $\epsilon = 0.56 \, \mathrm{kpc}$ and $N_\mathrm{ngb} = 64$ is used.
In Appendix~\ref{app:stability}, we demonstrate the stability of our initial conditions for a resolution of $N = 10^5$ particles when evolved without self-interactions. 

\begin{figure}
    \centering
    \includegraphics[width=\columnwidth]{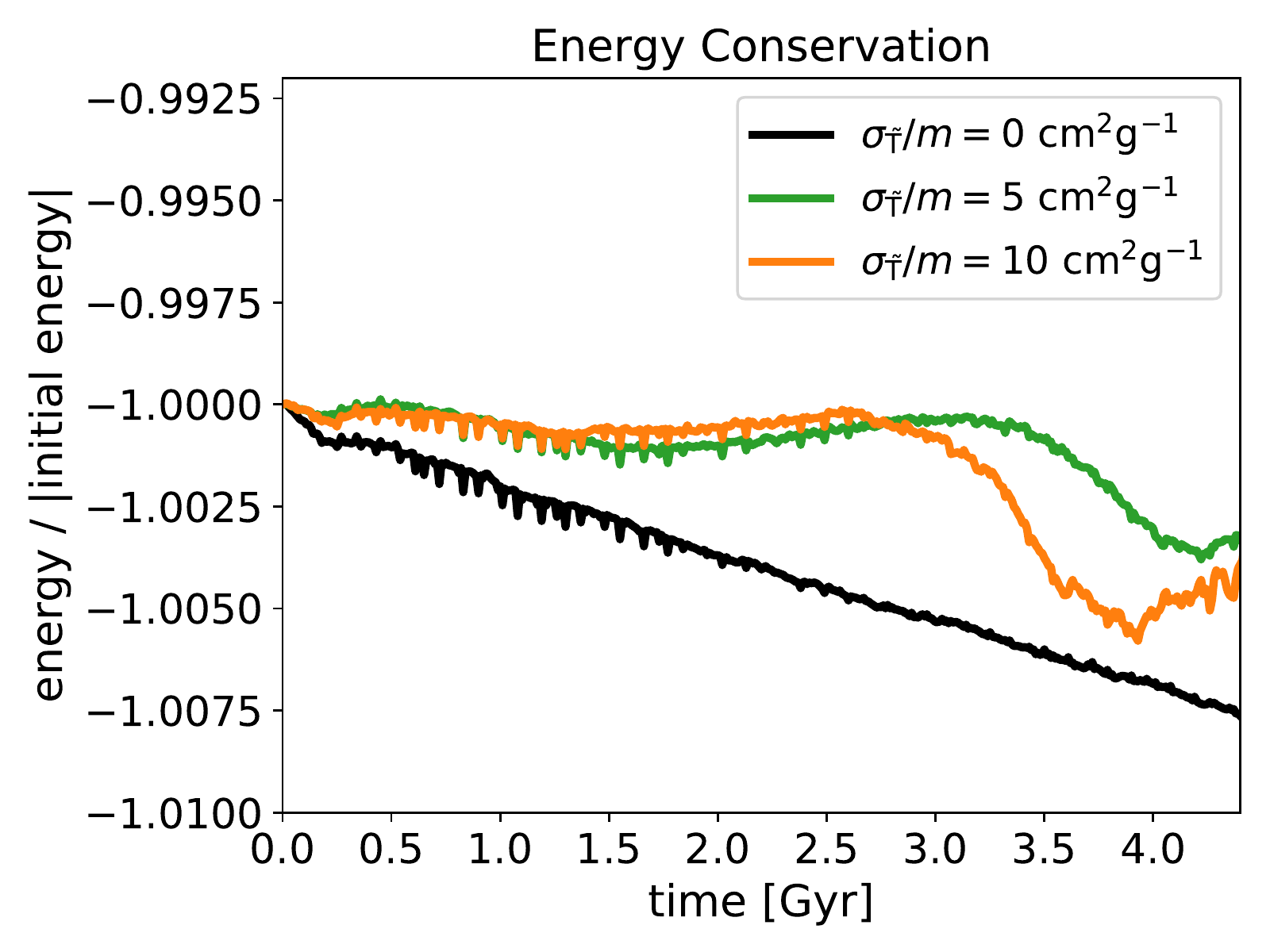}
    \caption{We show the evolution of the total energy for simulations of an initial NFW halo evolved with three different cross-sections. The black curve corresponds to $\sigma_\mathrm{\tilde{T}}/m = 0\, \mathrm{cm}^2\,\mathrm{g}^{-1}$, which is identical to the collisionless CDM.}
    \label{fig:halo_energy}
\end{figure}

First, we study the energy conservation for three different cross-sections using a resolution of $N=10^5$ particles.
For this purpose we compute the total energy of the entire halo and divide it by the absolute value of the initial total energy.
Our results are shown in Fig.~\ref{fig:halo_energy}.
The total energy is not perfectly conserved as the formulation of the Poisson solver does not explicitly conserve energy.
This is in contrast to the formulation of frequent self-interactions, which does conserve energy explicitly.
Nevertheless, the deviation from the initial energy is small enough that we can consider it as conserved for our purpose of astrophysical simulations.

\begin{figure}
    \centering
    \includegraphics[width=\columnwidth]{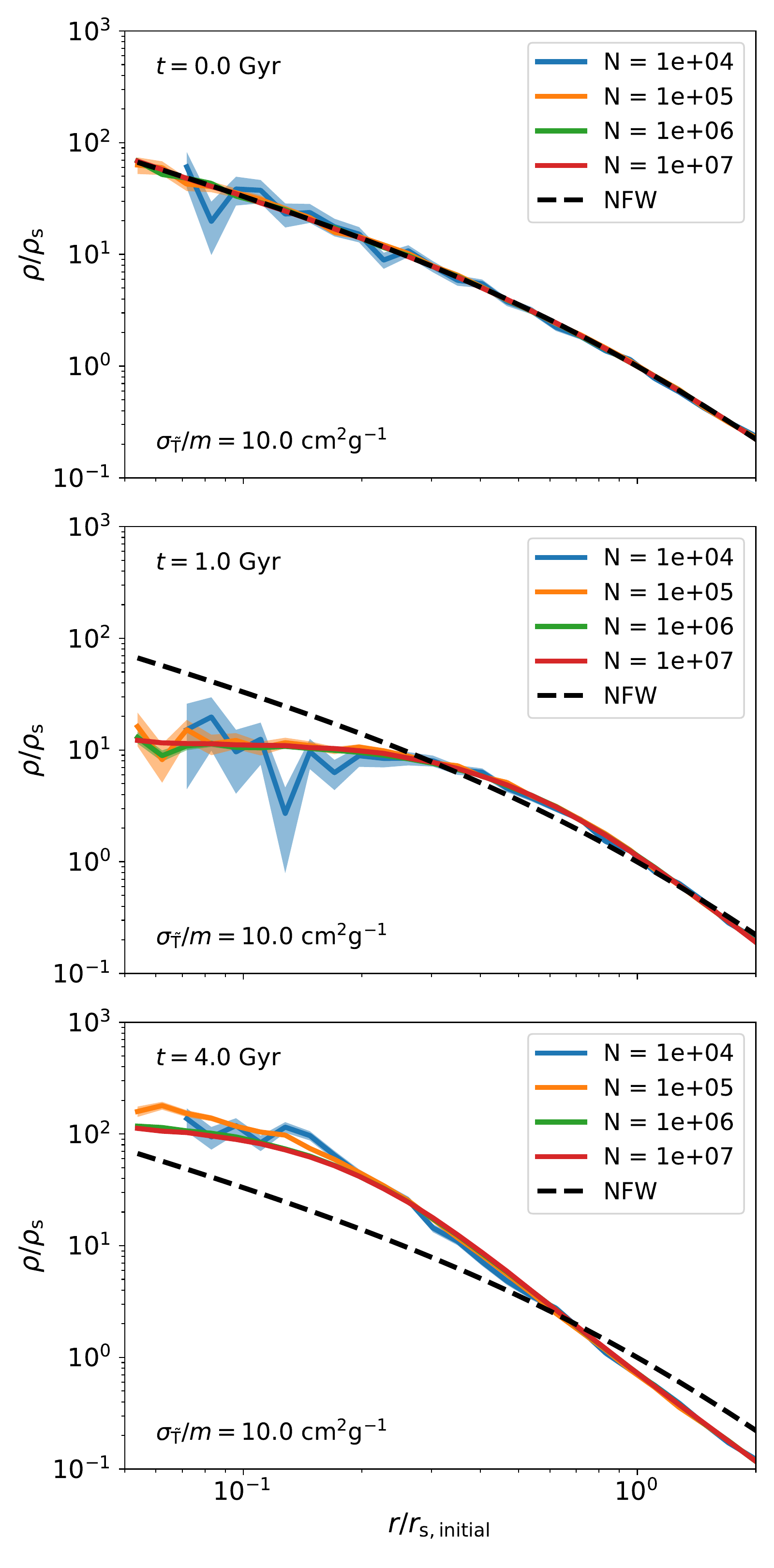}
    \caption{Density profiles for an initial NFW halo simulated with a cross-section of $\sigma_\mathrm{\tilde{T}}/m = 10\, \mathrm{cm}^2\,\mathrm{g}^{-1}$ are shown. Different resolutions were chosen to demonstrate convergence. The upper panel gives the initial conditions, the middle panel gives the DM halo after $1.0 \, \mathrm{Gyr}$ and the lower panel for $4.0 \, \mathrm{Gyr}$. For comparison, we show the analytical NFW profile in black.}
    \label{fig:halo_conv}
\end{figure}

Finally, we investigate the convergence of our numerical scheme.
We simulate the DM halo choosing different resolutions and a self-interaction cross-section of $\sigma_\mathrm{\tilde{T}}/m = 10\, \mathrm{cm}^2\,\mathrm{g}^{-1}$.
In Fig.~\ref{fig:halo_conv}, we show our results, i.e.\ density profiles at several times for different resolutions.
The upper panel represents the initial conditions, 
the middle and lower panel are for $t=1\,\mathrm{Gyr}$ and $t=4\,\mathrm{Gyr}$.
We can see that the density profile converges for increasing resolution, which confirms that our scheme is suitable for the application to astrophysical problems.

Fig.~\ref{fig:halo_conv} shows that initially a constant-density core forms, whereas at later times the central density increases.
This is because the self-interactions lead to a transport of energy in the outward direction. 
This energy loss causes the core to shrink which, eventually, leads to a gravothermal core-collapse like in rSIDM \citep[e.g.][]{Burkert_2000, Kochanek_2000, Koda_2011}. This process will be investigated in more detail in the following section.

\section{Core Size of Dark Matter Haloes} \label{sec:core-size}
In this section, we study the formation and evolution of a DM core in an isolated halo and compare frequent and rare self-interactions. We first describe the simulation set-up, then explain how we measure the core size and finally present our results.

As initial conditions, we take similar ones to \cite{Robertson_2017b}.
The initial density follows a Hernquist profile \citep{Hernquist_1990} with a mass of $M = 2.46 \times 10^{14} \, \mathrm{M_\odot}$ and a scale radius of $r_s = 279 \, \mathrm{kpc}$.
We sample the halo up to $r = 400 r_s$ using $N=10^7$ DM particles.

We explore the same cross-sections as in fig.~1 in \cite{Robertson_2017b}, chosen as $\hat{\sigma} \in \lbrace0, 1, 3, 10\rbrace$ with $\hat{\sigma} = (2\sigma_\mathrm{\tilde{T}}/m) (M/r_s^2)$, which implies $\sigma_\mathrm{\tilde{T}} / m \in \lbrace0, 0.227, 0.757, 2.272, 7.574 \rbrace \, (\mathrm{cm}^2 \, \mathrm{g}^{-1})$.\footnote{We use $\sigma_\mathrm{\tilde{T}}$ as defined in \cite{Kahlhoefer_2014}, which is a factor of 2 smaller in the regime of isotropic scattering compared to the commonly used values given in terms of $\sigma_\mathrm{T}$.}
We simulate these cross-sections both as frequent and rare scattering using our respective implementation in \textsc{gadget-3}, assuming the rare scattering to be isotropic.
This approach allows us to study differences between fSIDM and rSIDM in the context of core formation.
We conduct our simulations with a gravitational softening length of $\epsilon = 0.56 \, \mathrm{kpc}$ and use $N_\mathrm{ngb} = 64$ for the scattering.

In order to measure core sizes, we follow \cite{Robertson_2017b}, i.e.\ we fit a cored Hernquist profile to the radial density distribution,
\begin{equation} \label{eq:cored_hernquist}
    \rho(r) = \frac{M}{2 \pi} \, \frac{r_s}{( r^\beta + r_\mathrm{core}^\beta)^{1/\beta}} \, \frac{1}{(r\vphantom{r_\mathrm{core}^\beta} + r_s )^3} .
\end{equation}
As free parameters we take the core radius $r_\mathrm{core}$, the scale radius $r_s$ and the mass $M$, while $\beta$ is kept fixed to $\beta=4$.
We then determine the number of particles $N_i$ in several radial bins with boundaries $r_i$ and $r_{i+1}$ and compare this number to the expected value $\lambda_i$ according to the cored density profile:
\begin{equation}
    \lambda_i = \frac{4\pi}{m} \int_{r_i}^{r_{i+1}} r^2 \rho(r) \, \mathrm{d}r \, ,
\end{equation}
where $m$ denotes the mass of a simulation particle.
To fit the density profile we maximize a likelihood based on Poisson statistics,
\begin{equation}
    \mathcal{L} = \prod_i \frac{\lambda_i^{N_i} \, e^{-\lambda_i}}{N_i!} \,.
\end{equation}

\begin{figure}
    \centering
    \includegraphics[width=\columnwidth]{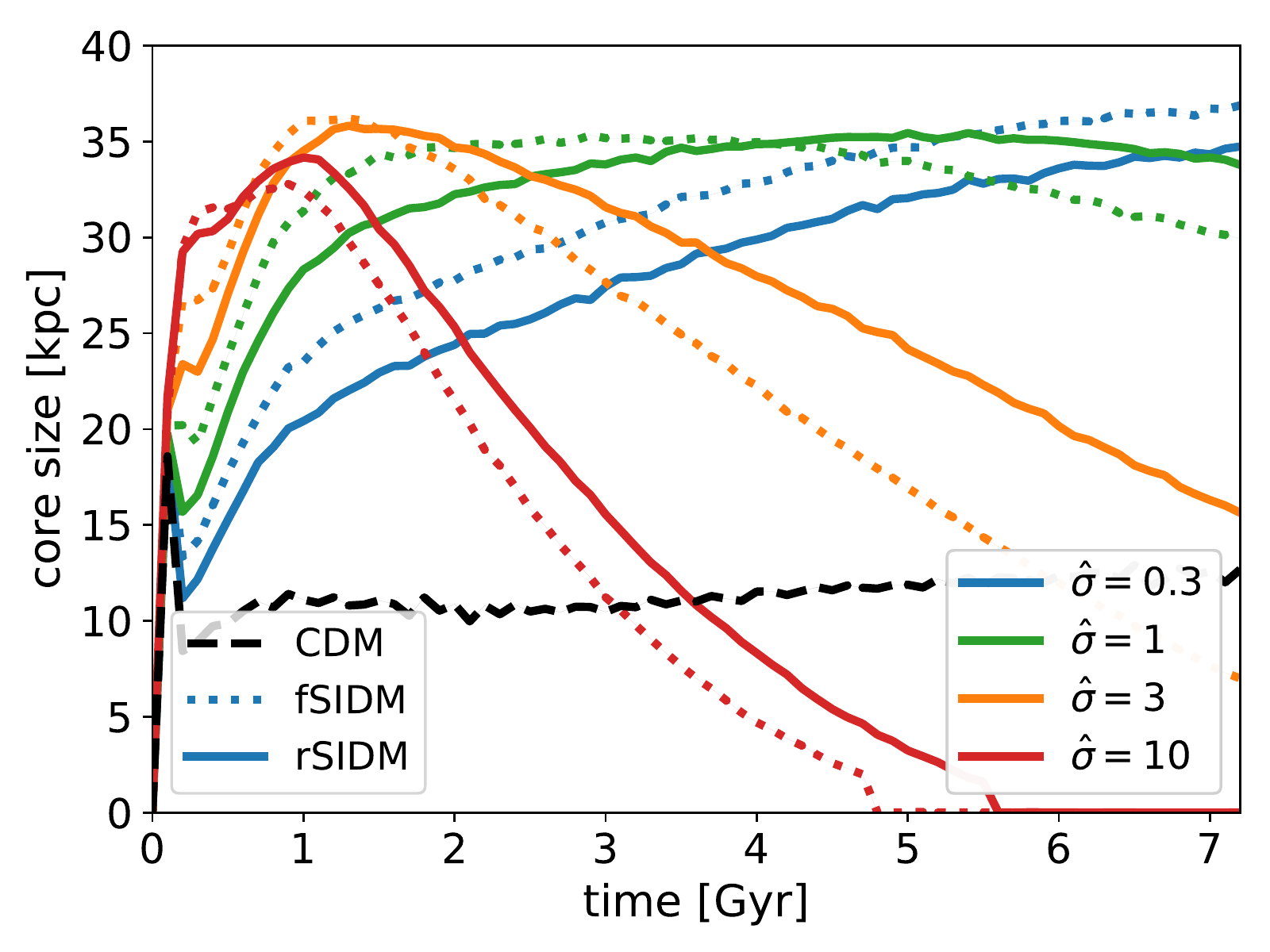}
    \caption{The evolution of the core size for an isolated DM halo is shown. The halo has initially the shape of an Hernquist profile. The halo was simulated using several cross-sections for frequent and rare self-interacting DM. The errors correspond to the $16\%$ and $84\%$ levels. They were computed using the Markov chain Monte Carlo sampling implementation of emcee \citep{emcee}.}
    \label{fig:core-size}
\end{figure}

In Fig.~\ref{fig:core-size}, we show the evolution of the core size over a time of $7.2 \, \mathrm{Gyr}$.
Within the studied range of $\Hat{\sigma}$, we find the time it takes for the core to grow and collapse to decrease with increasing self-interaction cross-section. Moreover, the core formation happens much faster than the core-collapse. We find that the maximum core size is nearly independent of the self-interaction cross-section for both rSIDM and fSIDM, in agreement with earlier findings for rare self-interactions \citep[e.g.][]{Kochanek_2000}.

When comparing frequent and rare self-interactions with the same momentum transfer cross-section, the evolution of fSIDM is a bit faster, i.e.\ maximum core size is reached earlier. Nevertheless, there is no big difference between frequent and rare scattering. The largest deviation is found at late times for a cross-section of $\hat{\sigma} = 3$. The other free parameters of our fit (e.g. $M$ and $r_s$) behave very similar for rSIDM and fSIDM. Similar to the difference in core size we observe differences in the central density of the halo.

In comparison to fig.~1 \cite{Robertson_2017b}, we find a smaller maximum core-size, but overall a similar evolution.
The differences may be due to slight differences in the initial conditions.
Here, we approximated the local velocity distribution of the halo by a Maxwellian to sample the initial velocities.
As one can see in Fig.~\ref{fig:core-size}, at the very beginning of the simulation a core forms. This is because the initial conditions are not in perfect equilibrium. Even for a CDM run with flawless initial conditions, the core formation cannot be avoided completely as numerical effects lead to a small core.

We also note that core-collapse happens much faster in isolated DM haloes than in cosmological simulations, where the core is heated up through late-time infall.

\section{Equal-Mass Merger} \label{sec:merger_sim}

In this section, we study the evolution of an equal-mass merger using frequent and isotropic rare scattering.
We investigate several cross-sections and compare the two types of scattering.
This is interesting because merging systems allow to constrain DM self-interactions.
The scattering does lead to drag-like behaviour under given circumstances.
This decelerates the DM component but does not affect the galaxies\footnote{Note that we treat galaxies as collisionless test particles in this work as is mostly done in the literature.}
and thus leads to an offset between the two.
There have been several studies on merging systems with DM self-interactions in the literature \citep[e.g.][]{Randall_2008, Kahlhoefer_2015, Robertson_2017a} as well as discussions on the size of observed offsets \citep[e.g.][]{Bradac_2008, Dawson_2012, Dawson_2013, Jee_2015, Harvey_2017, Peel_2017, Taylor_2017, Wittman_2018}.
There is also an extensive literature on how the self-interactions affect the merger evolution and under which conditions the picture of a drag force is appropriate \citep[e.g.][]{Markevitch_2004, Harvey_2014, Kahlhoefer_2014, Kim_2017b, Robertson_2017b}.
As the drag-like behaviour is expected to depend on the shape of the differential cross-section, merging systems potentially allow for constraining not only the strength of the self-interactions but also its angular dependence.

\begin{figure}
    \centering
    \includegraphics[width=0.99\columnwidth]{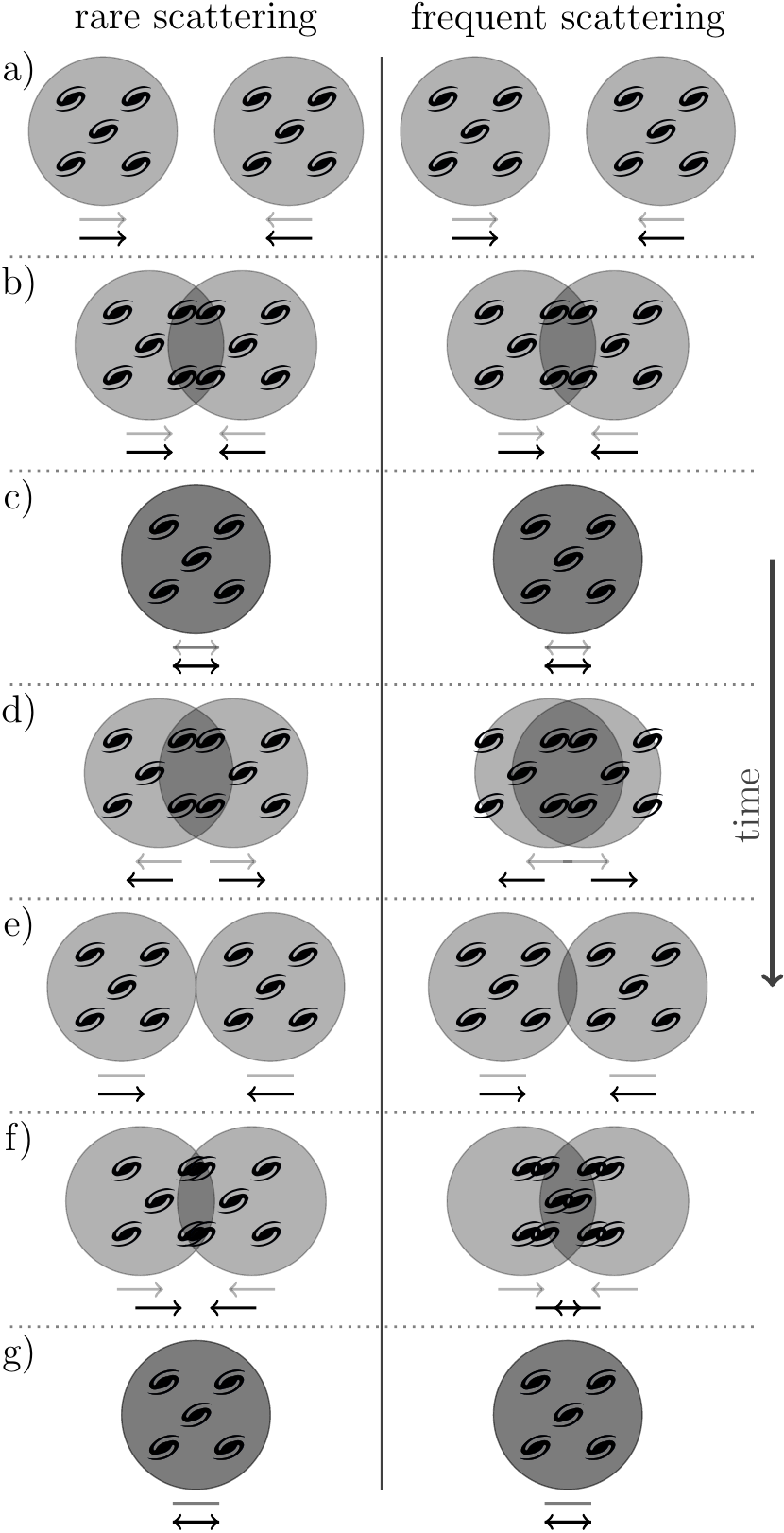}
    \caption{
    The evolution of an equal-mass galaxy cluster merger for frequent and rare DM self-interactions is illustrated. 
    We only illustrate the DM (grey circles) and Galaxy (black spirals) positions as well as their direction of motion indicated by the arrows.
    The shape of the DM haloes is not taken into account.
    Rare scattering is shown on the left-hand side and frequent scattering on the right-hand side.
    The time propagates from the top to the bottom.
    The evolution we illustrate here is similar to the one we found for a cross-section of $\sigma_\mathrm{\tilde{T}}/m = 1.5 \, \mathrm{cm}^2 \, \mathrm{g}^{-1}$, but exaggerated.
    In (a) we show the initial state and in (b) we illustrate the infall-phase.
    The first pericentre passage is displayed in (c) and (d) gives a time a little bit later.
    This is the first time where we find a significant difference between rSIDM and fSIDM.
    For the frequent interactions, the DM is closer to barycentre, but the galaxies behave similarly implying larger offsets for fSIDM.
    About the first apocentre passage both components reach a larger distance from barycentre if the self-interactions are rare.
    This is illustrated in (e).
    In (f), at a later time we find larger offsets for fSIDM, although the DM component is closer to barycentre than in rSIDM.
    Finally, we illustrate the second pericentre in (g).
    }
\label{fig:merger_cartoon}
\end{figure}

We start with a description of our simulation set-up and then explain how we analyse the simulation. 
Finally, we present and interpret our findings, in particular how the merger leads to offsets between DM and galaxies.
A schematic illustration of the merger is shown in Fig.~\ref{fig:merger_cartoon}. The various details shown in this figure will be discussed in the remainder of this section.

Our initial conditions are chosen similar to the ones of \cite{Kim_2017b}.
We set up two NFW haloes, each with a virial mass of $M_\mathrm{vir} = 10^{15}\,\mathrm{M_\odot}$. They are separated by $4000 \, \mathrm{kpc}$ and move initially with a relative velocity of $1000 \, \mathrm{km}\,\mathrm{s}^{-1}$ along the merger axis towards each other, such that the impact parameter of the merger is zero.
The two DM haloes are described by the same parameters but sampled independently.
The concentration parameter is $c=3.3$ and the scale radius is $r_\mathrm{s} = 630 \, \mathrm{kpc}$.
We sample each halo up to a radius of $2667 \, \mathrm{kpc}$ using $6 \times 10^6$ DM particles for each halo with a particle mass of $m_\mathrm{DM} = 2 \times 10^{8}\, \mathrm{M_\odot}$.

In addition, we include particles representing galaxies in our simulations.
Each halo has $3 \times 10^4$ of these particles with a mass of $m_\mathrm{Gal} = 8 \times 10^{8}\, \mathrm{M_\odot}$ each.
These particles do not represent individual galaxies (they are more abundant than galaxies in clusters) but they can be seen as a ``smoothed out'' galaxy distribution.
As in \cite{Kim_2017b}, we place a particle at the centre of each halo to model the brightest cluster galaxy (BCG).
These particles have a mass of $m_\mathrm{BCG} = 7 \cdot 10^{10} \, \mathrm{M_\odot}$.
This is a very idealized treatment of the BCGs as we neglect their extension.
In Appendix~\ref{app:stability}, we demonstrate that the haloes used for the merger simulation are stable when simulated in isolation without self-scattering.

We simulate the same self-interaction cross-sections as in \cite{Kim_2017b} plus some additional ones, i.e.\ $\sigma_\mathrm{\tilde{T}}/m \in \lbrace0.0, 0.5, 1.0, 1.5, 2.0, 2.5, 3.5, 5.0 \rbrace (\mathrm{cm}^2 \, \mathrm{g}^{-1})$.
In practice we match rare and frequent cross-section using $\sigma_\mathrm{\tilde{T}} = \sigma/2$ as appropriate for isotropic scattering (see Appendix~\ref{sec:rSIDM_implementation}).\footnote{This definition differs from the one in \cite{Kim_2017b}, where $\sigma_\mathrm{T} = \sigma$ is used.}
For the gravitational softening length we use a value of $\epsilon = 0.56 \, \mathrm{kpc}$ and employ  $N_\mathrm{ngb} = 64$ for the scattering.

\subsection{Method of analysis}

Before discussing our results in detail, let us first give an overview of the various figures that we have produced and the methods used to obtain them.
To analyse the simulations we find the peaks of the DM and galactic component (see Fig.~\ref{fig:merger_byc_dist2}).
Several methods for peak finding can be found in the literature.
In this work, we follow the algorithm described in \cite{Kim_2017b}, i.e.\ we use a kernel density estimate (KDE) with a 2D Gaussian smoothing kernel with a width of $100 \, \mathrm{kpc}$, while we project along one axis perpendicular to the merger axis.
As we only study simulations with an impact parameter equal to zero, we perform the peak search only along the merger axis, i.e.\ we take the positions with maximum density according to the KDE.
In order to obtain uncertainties on the peak position, we bootstrap the galaxy distribution 1000 times and the much better sampled DM component 10 times.

We then compute offsets between the components as the distance between their density peaks.
Therefore we define	the ``half''-separation between two peaks of the same species, i.e.\ the two DM peaks, the two galaxy peaks or the two BCGs.
\begin{equation}
	d := \frac{|x_1-x_0|}{2} \, ,
\end{equation}
where x is the x coordinate with respect to barycentre.
The offsets shown in Fig.~\ref{fig:merger_offsets} are the mean offset of the two haloes, with positive values implying that the galaxies are closer to the barycentre than the DM and negative values corresponding to the opposite case.
In Fig.~\ref{fig:merger_maxoffsets}, we show the maximum positive offset as function of the self-interaction cross-section.
Furthermore, we compute how much the peaks for fSIDM and rSIDM deviate from each other (Fig.~\ref{fig:merger_peak_dev}).
For this purpose, we define a quantity $\delta$ that is based on the mean of the two haloes,
\begin{equation} \label{eq:peak_dev}
    \delta := \frac{|x_{\mathrm{r},0}-x_{\mathrm{r},1}|-|x_{\mathrm{f},0}-x_{\mathrm{f},1}|}{2} \, .
\end{equation}
Here, $x$ denotes the peak position of the DM haloes (0 and 1) for rare (r) and frequent (f) self-interactions.
A positive value of $\delta$ implies that the fSIDM peaks are closer to barycentre than the rSIDM peaks and vice versa.

When the peak separation is small the peak identification becomes inaccurate and biased towards the barycentre (as can be seen in Fig.~\ref{fig:merger_byc_dist2}).
This is why we do not show offsets and peak deviation for separations less than the scale radius ($r_s = 630 \, \mathrm{kpc}$).
We also do not consider these values for the subsequent analysis.

\subsection{Results}
\label{sec:merger_findings}
\begin{figure}
    \centering
    \includegraphics[width=\columnwidth]{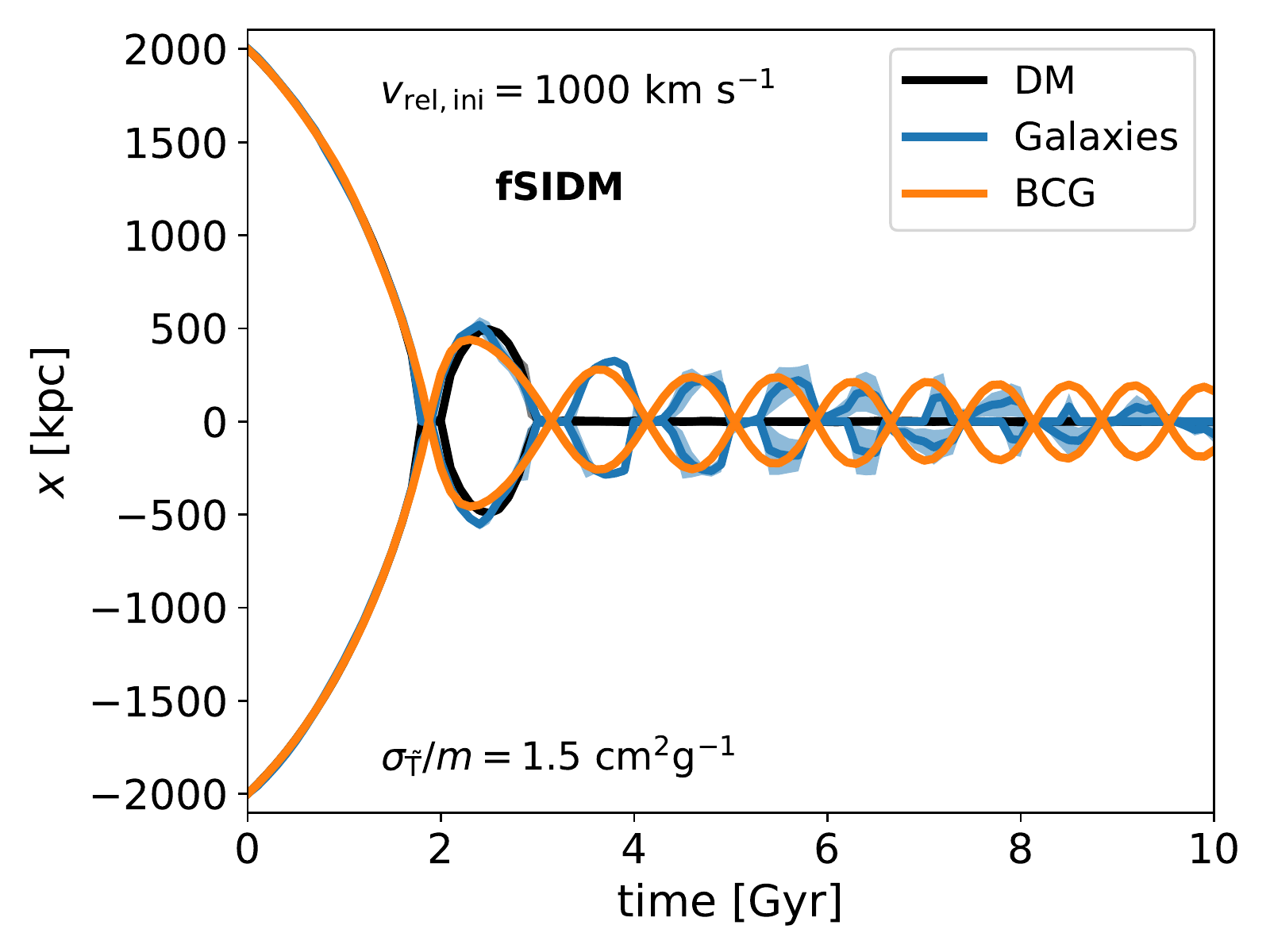}
    \includegraphics[width=\columnwidth]{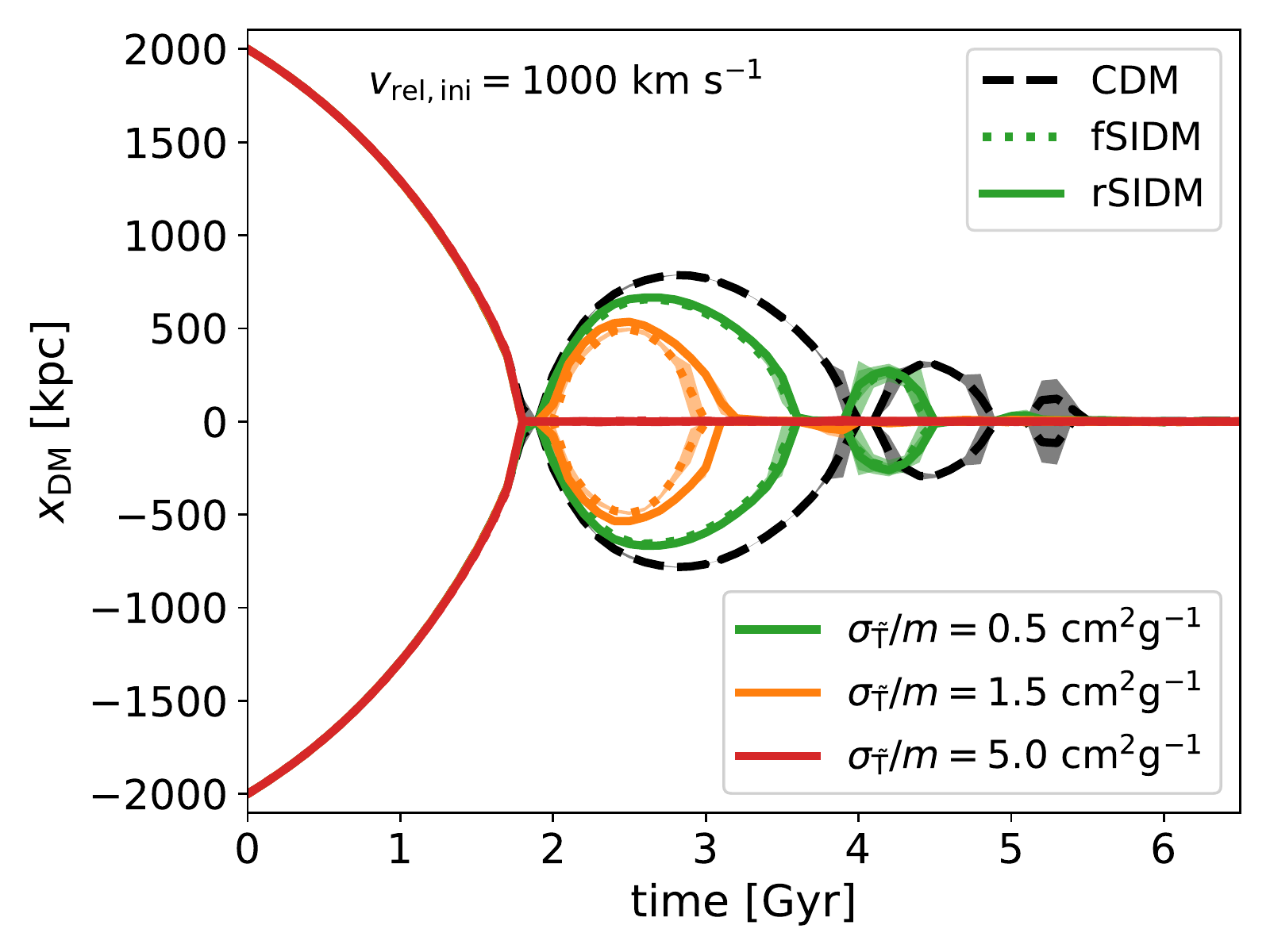}
    \caption{Upper panel: The density peak distance to barycentre for various components of a merger is shown as a function of time.
    Two NFW haloes were merged using frequent self-interacting DM with a cross-section of $\sigma_\mathrm{\tilde{T}}/m = 1.5\, \mathrm{cm}^2 \, \mathrm{g}^{-1}$.
    We measure the density peak for each of the two haloes.
    We do this separately for the DM and galaxies. Each halo contains one particle to model the BCGs.
    For the plot we simply use the position of that particle.
    The plot shows the distance to the barycentre along the merger axis.
    Lower panel: The plot is similar to the upper one.
    Here, we show the DM component only, but for several merger simulations with different self-interaction cross-sections.}
    \label{fig:merger_byc_dist2}
\end{figure}

The upper panel of Fig.~\ref{fig:merger_byc_dist2} shows how the density peaks of all components evolve with time for frequent scatterings with a cross-section $\sigma_\mathrm{\tilde{T}}/m = 1.5 \, \mathrm{cm}^2 \, \mathrm{g}^{-1}$.
The DM component coalesces earlier than the galactic component due to the self-interactions.
Similar to \cite{Kim_2017b}, we find long-lasting oscillations of the BCG particles.
The same plot for several other runs can be found in Appendix~\ref{ap:merger_plots}.
From these plots, we can see that the galaxies and BCGs behave differently, depending on the type of DM self-interaction.
An exaggerated illustration of the merger evolution for rSIDM and fSIDM inspired by simulations with a cross-section of $\sigma_\mathrm{\tilde{T}}/m = 1.5 \, \mathrm{cm}^2 \, \mathrm{g}^{-1}$ is shown in Fig.~\ref{fig:merger_cartoon}.

In general, we find that larger cross-sections lead to shorter merger times for both rSIDM and fSIDM.
This is shown in the lower panel of Fig.~\ref{fig:merger_byc_dist2}, where we show the evolution of the DM peak position for selected merger simulations.
Furthermore, the distance of the DM peaks at first apocentre passage shrinks with increasing cross-section.
The evolution of the DM peaks for rSIDM and fSIDM is similar but not identical.
For the shown simulations the largest difference occurs in our run with a self-interaction cross-section of $\sigma_\mathrm{\tilde{T}}/m = 1.5 \, \mathrm{cm}^2 \, \mathrm{g}^{-1}$.
For large cross-sections the differences vanish since the two haloes coalesce on contact.

\begin{figure*}
    \centering
    \includegraphics[width=\textwidth]{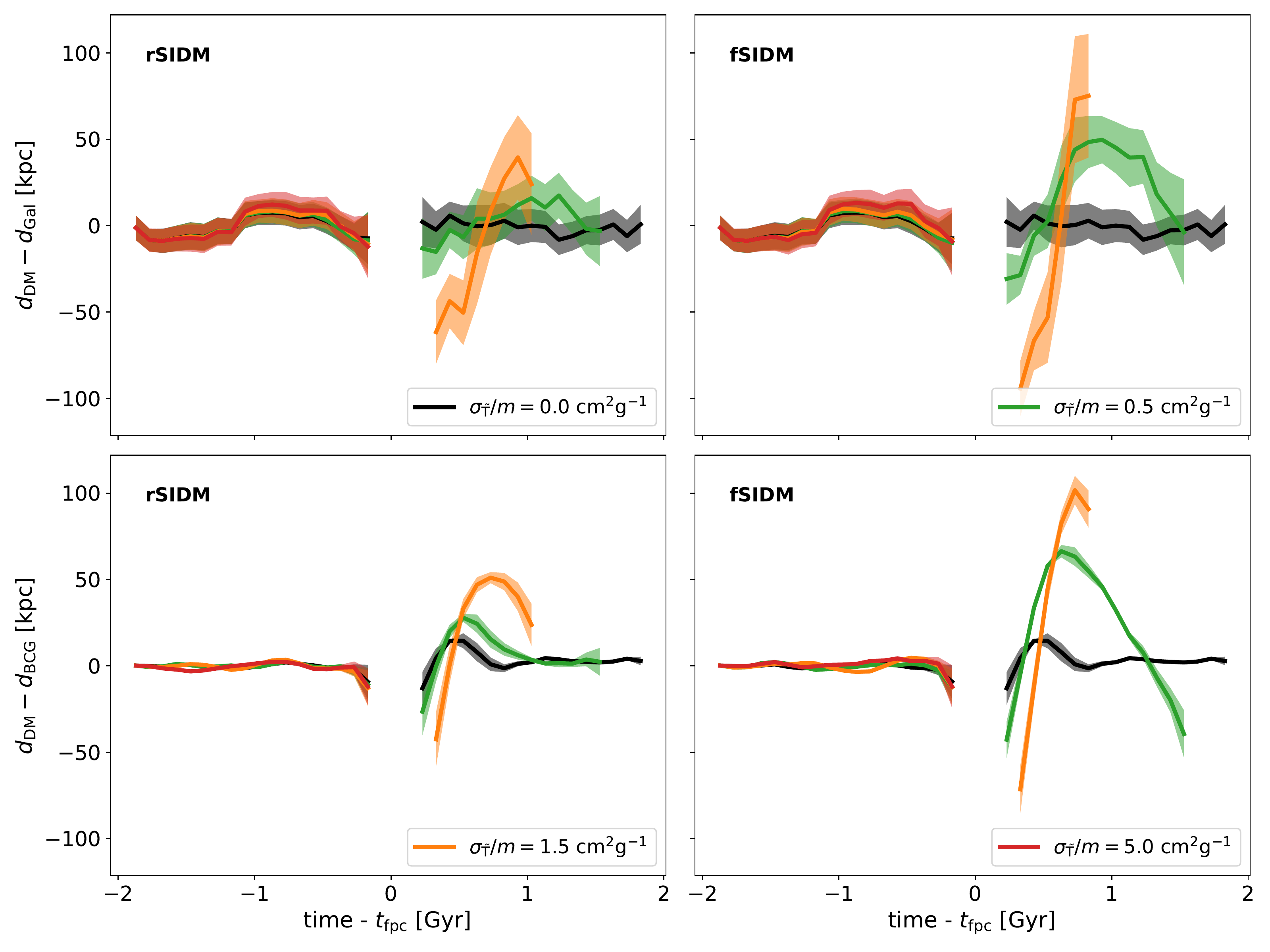}
    \caption{Offsets between DM and galaxies (upper panels) or BCGs (lower panels) are shown as function of time.
    Here we measure the time with respect to the first pericentre passage ($t_\mathrm{fpc} = 1.87 \, \mathrm{Gyr}$).
    We display results for several self-interaction cross-sections.
    The left-hand panels give the offsets for rare self-interactions and the right-hand panels for frequent scattering.
    The galaxy offsets before the first pericentre passage are mainly due to the uncertainty in the galaxy peaks (compare upper and lower panels).}
    \label{fig:merger_offsets}
\end{figure*}

Next, we study galaxy--DM and BCG--DM offsets for fSIDM and rSIDM, which are shown in Fig.~\ref{fig:merger_offsets}. 
In general, we find the offsets to be larger for fSIDM when comparing to the same rSIDM momentum transfer cross-section.
Also, the offsets of the BCG particles are larger than the offsets of the galactic component.
This is probably a consequence of modelling them as point-like instead of treating them as extended objects.
For $\sigma_\mathrm{\tilde{T}}/m = 1.5 \, \mathrm{cm}^2 \, \mathrm{g}^{-1}$, the offsets are zero when the galaxies are roughly at apocentre, but before and afterwards they are non-zero with different signs.
Compared to the first apocentre passage the point in time when the sign of the offsets changes becomes earlier with decreasing cross-section.
For the early offsets, the DM component is closer to barycentre (i.e.\ the offset is negative), but for the late offsets, the DM is more distant from the barycentre than the galaxies (i.e.\ the offset becomes positive).
It is worth mentioning that the difference between fSIDM and rSIDM in the offsets shortly after the first pericentre passage are mainly due to different peak positions of the DM as we explain below.
But for the later offsets it is the other way around because then the offsets are caused by differences in the galaxy peak positions.
Note, here we only considered offsets before the second pericentre passage. For even later offsets, the sign potentially changes again, but typically the offsets are smaller.

\begin{figure}
    \centering
    \includegraphics[width=\columnwidth]{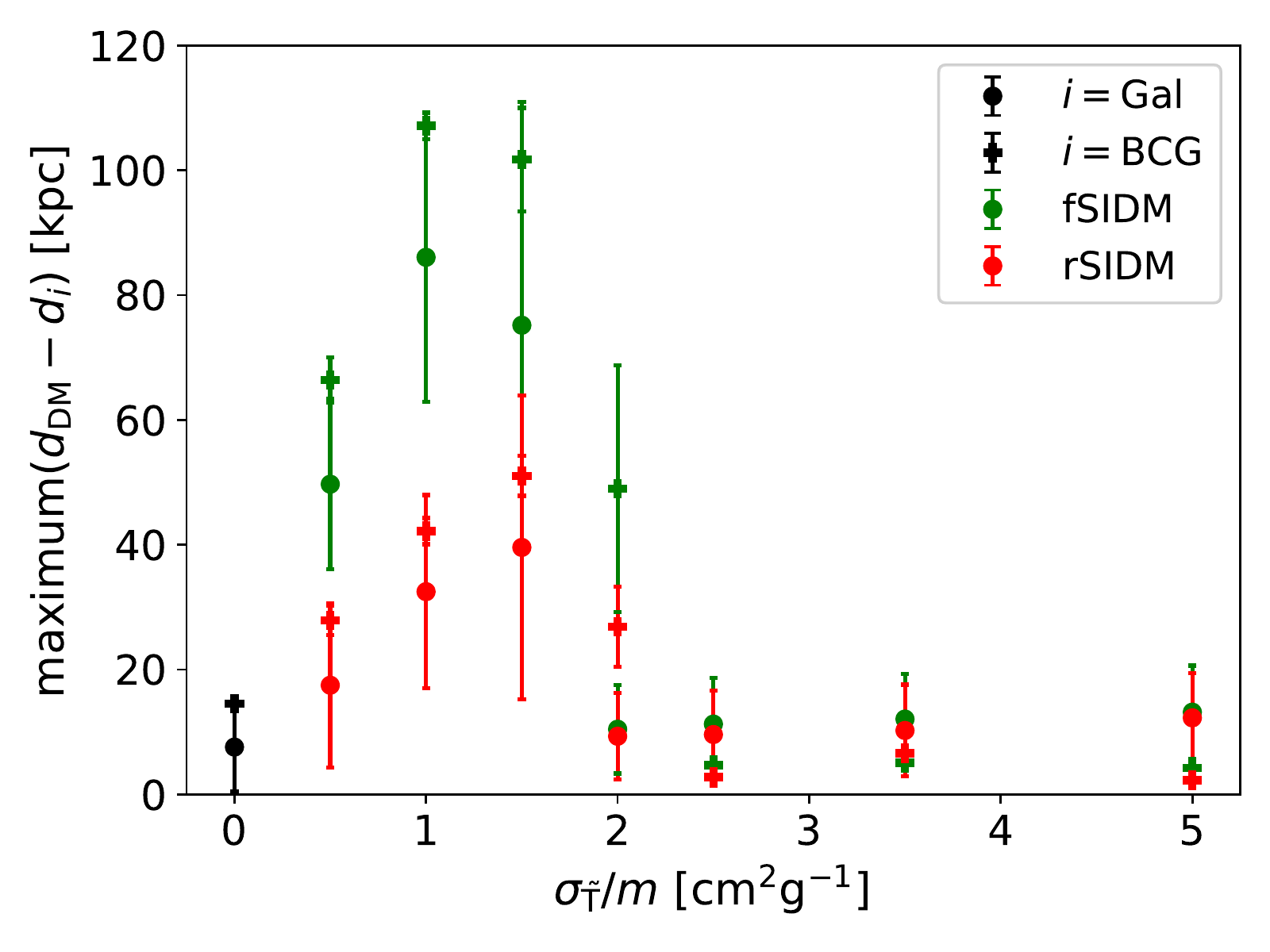}
    \caption{We show the maximum offset as function of self-interaction cross-section.
    We consider the distance between DM peaks and the peak of the galactic component or the BCG as shown in Fig.~\ref{fig:merger_offsets}.
    The offsets are shown for both, frequent and rare self-interactions.
    It should be mentioned that we only consider offsets where the DM component is more distant to the centre of mass than the galaxies.
    The shown results for $\sigma_\mathrm{\tilde{T}}/m \in \lbrace2.0, 2.5, 3.5\rbrace (\mathrm{cm}^2 \, \mathrm{g}^{-1})$ are likely inaccurate due to the peak finding method.}
    \label{fig:merger_maxoffsets}
\end{figure}

In the literature, the early offsets, e.g.\ for the Bullet Cluster, have been studied, which arise directly after the first pericentre passage.
In contrast, we will mainly focus on a later stage of the merger evolution.
In Fig.~\ref{fig:merger_maxoffsets}, we compare the maximum size of the offsets in the stage where the galaxies are closer to barycentre.
We find the largest offset for fSIDM in the  simulation with $\sigma_\mathrm{\tilde{T}}/m = 1.0 \, \mathrm{cm}^2 \, \mathrm{g}^{-1}$ and for rSIDM in the simulation with $\sigma_\mathrm{\tilde{T}}/m = 1.5 \, \mathrm{cm}^2 \, \mathrm{g}^{-1}$.
The largest fSIDM offset is more than a factor of 2 larger than the largest rSIDM offset.
In other words, frequent self-interactions can cause much larger offsets (when the galaxies are closer to barycentre) than rare self-interactions.
For smaller cross-sections ($\sigma_\mathrm{\tilde{T}}/m \lesssim 1.0 \, \mathrm{cm}^2 \, \mathrm{g}^{-1}$), the maximum offset decreases, but there are still difference of more than a factor of 2 between fSIDM and rSIDM.
The larger offsets of fSIDM at small cross-sections ($\sigma_\mathrm{\tilde{T}}/m \sim 0.5 \, \mathrm{cm}^2 \, \mathrm{g}^{-1}$) are particularly interesting as they could potentially be observable.

For large cross-sections ($\sigma_\mathrm{\tilde{T}}/m \gtrsim 1.5 \, \mathrm{cm}^2 \, \mathrm{g}^{-1}$), the maximum offsets decrease with increasing cross-section and so does the difference between simulations of rare and frequent scattering.
For $\sigma_\mathrm{\tilde{T}}/m \gtrsim 5.0 \, \mathrm{cm}^2 \, \mathrm{g}^{-1}$ the DM haloes coalescence on contact and the type of offsets we discuss here no longer occurs.
We note that measuring offsets with our peak finding method could be inaccurate for some cross-sections, i.e.\ for $\sigma_\mathrm{\tilde{T}}/m \in \lbrace2.0, 2.5, 3.5\rbrace (\mathrm{cm}^2 \, \mathrm{g}^{-1})$, since we neglect the offsets for small halo separations as the peaks are biased towards barycentre.

\begin{figure}
    \centering
    \includegraphics[width=\columnwidth]{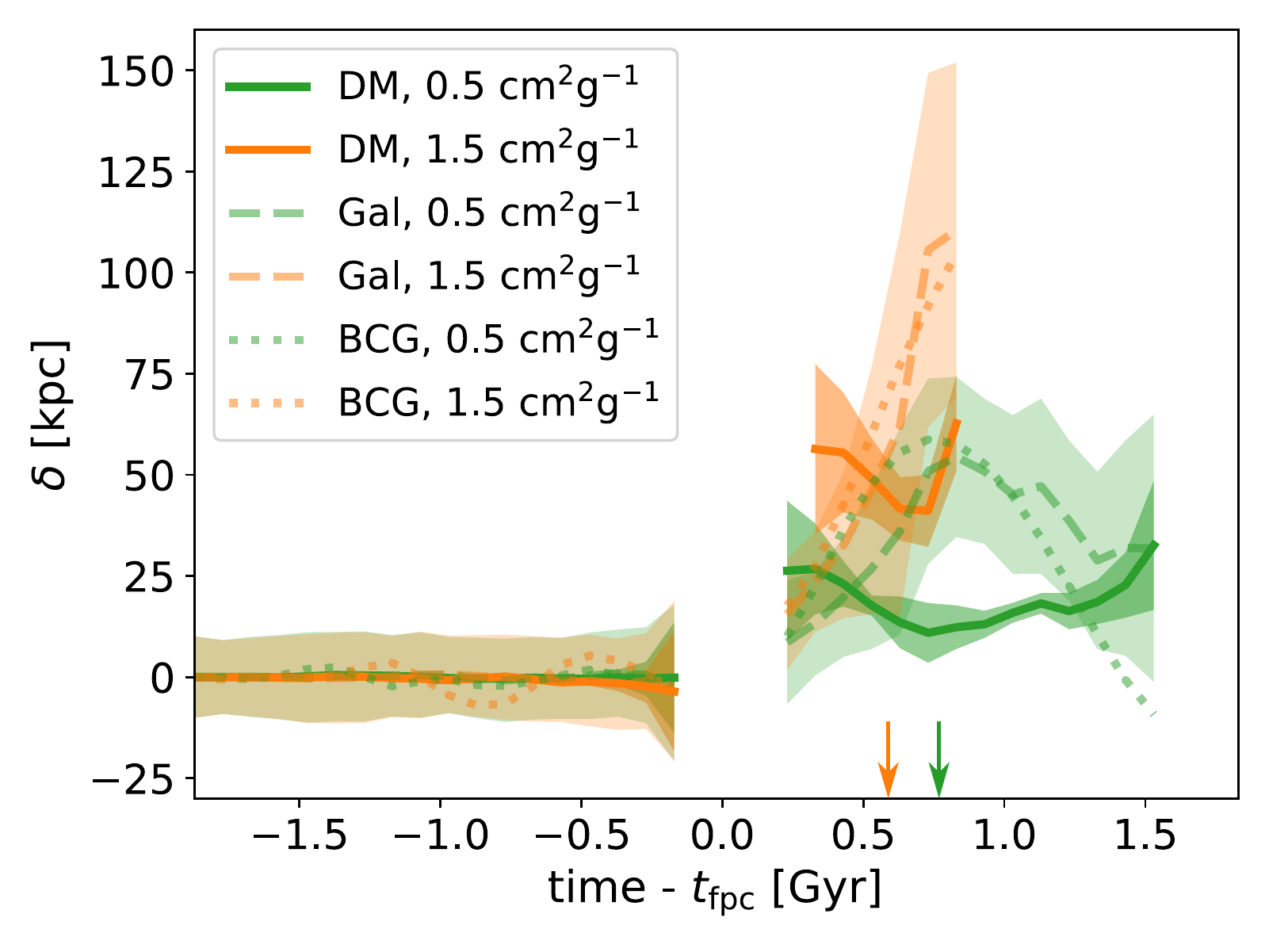}
    \caption{The deviation ($\delta$, see equation~\ref{eq:peak_dev}) of peaks between the fSIDM and rSIDM runs is shown as function of time.
    We measure the time with respect to the first pericentre passage ($t_\mathrm{fpc} = 1.87 \, \mathrm{Gyr}$).
    A positive value of $\delta$ implies that the peak of the fSIDM simulation is closer to barycentre than the rSIDM one.
    We compare DM and galaxy peaks as well as the positions of the BCGs.
    Results are plotted for $\sigma_\mathrm{\tilde{T}}/m = 0.5 \, \mathrm{cm}^2 \, \mathrm{g}^{-1}$ (green) and $\sigma_\mathrm{\tilde{T}}/m = 1.5 \, \mathrm{cm}^2 \, \mathrm{g}^{-1}$ (orange).
    Note, the peak deviation is only shown when the distance of the peaks is larger than the scale radius ($630 \, \mathrm{kpc}$).
    We also apply this to the BCGs.
    The first apocentre passage (which is very similar for rSIDM and fSIDM) is indicated by an arrow for each cross-section.}
    \label{fig:merger_peak_dev}
\end{figure}

Finally, we compare the peak positions in rSIDM and fSIDM.
In Fig.~\ref{fig:merger_peak_dev} we show the quantity $\delta$ defined in equation~\ref{eq:peak_dev} and find that $\delta$ increases with $\sigma_\mathrm{\tilde{T}}/m$ in the regime of small cross-sections.
Of particular interest is the evolution between the first pericentre passage and the second one, which occurs $\sim 1.9 \, \mathrm{Gyr}$ after the first one for $\sigma_\mathrm{\tilde{T}} / m = 0.5 \, \mathrm{cm}^2\,\mathrm{g}^{-1}$ and for larger cross-sections earlier. 
The DM peaks of the fSIDM run are found to be closer to the barycentre than for the corresponding rSIDM run, corresponding to $\delta > 0$ (solid lines).
The same is true for the galaxies (dashed lines) and the BCGs (dotted lines).
Shortly after the pericentre passage, $\delta$ is smaller for the galaxies than for the DM component because only the DM and not the galaxies are affected by the self-interactions.
However, the deviation of the galaxy peaks grows subsequently and becomes larger than the one for DM well before the first apocentre ($\sigma_\mathrm{\tilde{T}}/m = 0.5 \, \mathrm{cm}^2 \, \mathrm{g}^{-1}$) or at a somewhat later time around the first apocentre  ($\sigma_\mathrm{\tilde{T}}/m = 1.5 \, \mathrm{cm}^2 \, \mathrm{g}^{-1}$).
This is a consequence of how the galaxies respond to differences in the DM distribution via gravitational interaction.
This response leads to a greater difference in the galaxy distribution and creates the larger offsets for frequent scattering compared to rare scattering shown in Fig.~\ref{fig:merger_maxoffsets}.
In Appendix~\ref{ap:merger_amplification}, we provide further details on this amplification mechanism.

Overall, we found that the phenomenology of fSIDM differs significantly from the one of rSIDM.
In particular, frequent self-interactions can lead to much larger offsets than rare scattering.
Consequently, it should be possible at least in principle to distinguish between the two types of DM self-interactions using detailed observations of merging galaxy clusters.

\section{Discussion} \label{sec:discussion}

In this section, we first discuss technical issues concerning the numerical scheme, its implementation and the analysis of our simulation.
Then we elaborate on the physical implications of our results.

\subsection{Technical aspects}

From a technical perspective, there are several interesting directions for future extensions and improvements. An obvious next step would be to include an angular dependence in the rSIDM scheme \citep{Robertson_2017b}. 
It should then be possible to simulate arbitrary differential cross-sections, including those that have significant scattering probabilities in both the rare and the frequent scattering regime.
For this purpose one could for example introduce a cut-off angle that distinguishes between the two cases, such that small-angle scattering is treated in the frequent regime while large-angle scattering is simulated explicitly in the scheme for rare scattering. To validate this approach one needs to confirm that results do not depend on the precise value of the cut-off angle.

Another important extension will be to model velocity-dependent differential cross-sections, which appear to be preferred by observational data \citep[e.g.][]{Correa_2021, Sagunski_2021} and have been investigated in several $N$-body studies \citep[e.g.][]{Colin_2002,Vogelsberger_2012,Banerjee_2020}.
In fact such a velocity dependence is very natural from the particle physics perspective, in particular for frequent DM scatterings induced by light-mediator exchange, see e.g.~\cite{Buckley:2009in,Loeb:2010gj, Bringmann:2016din}. 
Such a velocity dependence can be easily implemented in our code, for both rSIDM as well as fSIDM.
Furthermore, one could abandon the assumption that scattering is elastic and also model dissipative scattering processes within the \textit{N}-body method \citep{Huo_2019}.

In addition to the scattering process, one could improve the treatment of the galactic component in our simulations. In this study we have treated galaxies as collisionless particles, which may be inaccurate to some extent \citep{Kummer_2018}. Moreover, galaxies are extended objects and their size may be too large to be approximated by point masses. Especially the trajectory of BCGs could be affected in relevant ways.  

As mentioned in the description of the implementation, the parallelization of the frequent scattering is presently not optimal.
One can envision a better scheme that exploits symmetries and does not cause large latency times, even though such a scheme could not make use of the infrastructure (e.g.\ tree algorithm, domain decomposition) that \textsc{gadget-3} provides.
However, for our purpose the current parallelization is sufficient to complete our simulations in reasonable times.
For example, consider the simulations presented in sec.~\ref{sec:merger_sim}, which were executed using MPI parallelization only on 64 logical cores.
The computation of the frequent scattering took $\sim 80 \%$ of the computing time, out of which a quarter was spent on the scattering itself. The rest of the time was used for other parts of the calculation, such as the neighbour search and the parallelization overhead. In comparison the scheme for rare self-interactions is less complicated and needs less computation time.

The robustness of our implementation could be increased by using an additional time-step constraint for the self-interactions. This would make the simulation code more capable of handling situations like gravothermal core-collapse of DM haloes. For the simulations we presented here, we only relied on the gravitational time-step, which is small enough for the situations we considered.

Finally, for our merger simulations, we used an algorithm to find peaks of the DM and the galactic component based on KDEs.
Unfortunately, the results are biased towards the barycentre for small peak separations, which limits the conclusion that can be drawn.
Other methods may perform better, for example finding the most tightly bound particle.

\subsection{Physical considerations}

In our various simulations, we found that fSIDM and rSIDM lead to different effects even when using the same momentum transfer cross-section. One may wonder whether this is simply the result of an incorrect matching, i.e.\ whether for each fSIDM cross-section one can find an rSIDM cross-section that produces the same behaviour. Indeed, Fig.~\ref{fig:core-size} suggests that core formation simply proceeds a bit faster in fSIDM than in rSIDM and it should be possible to improve the matching by using slightly larger cross-sections for rare scattering. However, given the physical difference between rare and frequent scattering it is also conceivable that the two cases cannot be matched to one another in a simple way. 
Rare self-interactions affect only a few particles per time, whereas frequent self-interactions affect all particles, which could contribute to a faster core-collapse for fSIDM.
One consequence of this is that systems with frequent or rare scatterings follow different paths to equilibrium, i.e. the velocity distributions are different. This can be seen from the idealized case studied in Fig.~\ref{fig:gadget_test_vel_dist2}.

Nevertheless, when we go beyond relaxed systems we find that the two types of self-interactions lead to qualitatively different effects. This can be seen most clearly in Fig.~\ref{fig:merger_maxoffsets}, where the largest offsets found in fSIDM cannot be reproduced for any cross section in rSIDM. In other words, frequent and rare self-interactions cannot be matched by a simple rescaling of the cross-section. It is worth mentioning that the cross-section needed to create the largest possible offset for a given system depends on the central density of the system. Thus, not all systems would allow rare and frequent scatterings to be distinguished observationally, and in the case of very small cross-sections all systems would be consistent with either rare or frequent scatterings. However, in the past, there have been claims of observations of large offsets \citep[][though see \citet{Wittman_2018}]{Harvey_2015}. Observed offsets are typically smaller ($\lesssim 60 \, \mathrm{kpc}$) than the one that can only be explained with frequent self-interactions in our simulations. Nevertheless, the observed offset of the Musket Ball Cluster ($\sim 80 \, \mathrm{kpc}$) is larger \citep{Dawson_2013}.

Furthermore, we have found that the difference between frequent and rare self-interactions results in an amplified difference in the galactic component, i.e.\ the maximum $\delta$ is larger in the galaxy distribution than for DM. Large offsets are easier to detect and their existence or non-existence has the potential to distinguish between frequent and rare scattering. Hence, the amplification process for fSIDM provides an important handle for determining the nature of DM.

Finally, we emphasize that we have adopted a simplified and idealized set-up in our simulations. For instance, we do not include baryons which could affect our results \citep{Zhang_2016}.
For a detailed comparison with observations, more realistic simulations will be required.
As mentioned above, such future simulations should also investigate in more detail the case of non-isotropic and velocity-dependent self-interactions, which has been found to have a significant impact on the offsets in merging galaxy clusters~\citep{Robertson_2017b} for rSIDM.

\section{Summary and Conclusions} \label{sec:conclusion}

In this paper, we have presented a novel method for modelling frequent self-interactions of DM within the framework of the \textit{N}-body method.
Our numerical scheme conserves energy and momentum explicitly.
Moreover, it does not rely on equilibrium or quasi-equilibrium states but is capable of treating typical astrophysical initial conditions.
We introduced several test problems to demonstrate the accuracy of our numerical scheme.
Furthermore, we performed several simulations of isolated haloes and mergers using frequent and rare self-interactions.
Our main results from these simulations are as follows:
\begin{itemize}
    \item fSIDM can be modelled accurately within \textit{N}-body simulations.

    \item Rare and frequent interactions lead to similar core formation in DM haloes. When considering the same momentum transfer cross-section, the evolution is slightly faster for fSIDM than for rSIDM.

    \item We found that fSIDM produces larger DM--galaxy offsets than rSIDM in equal-mass mergers.
        
    \item This effect can be traced back to an amplification in the displacement of the galactic component for the case of fSIDM. 

    \item In conclusion, the phenomenology of rSIDM and fSIDM is different in the sense that for a given strength of frequent self-interactions one cannot in general find a rare self-interaction cross-section that gives the same effects. 
\end{itemize}

This paper only constitutes the first steps towards exploring the astrophysical phenomenology of frequently self-interacting DM, in the sense that it provides the numerical methods for further investigations.
Future simulations of various astrophysical set-ups may provide deeper insights into the phenomenology of fSIDM and allow for a detailed comparison with observations.

\section*{Acknowledgements}
This work is funded by the Deutsche Forschungsgemeinschaft (DFG, German Research Foundation) under Germany's Excellence Strategy -- EXC 2121 ``Quantum Universe'' --  390833306, Germany’s Excellence Strategy -- EXC-2094 ``Origins'' -- 390783311 and the Emmy Noether Grant No.\ KA 4662/1-1. AR is supported by the European Research Council's Horizon2020 project `EWC' (award AMD-776247- 6). Preprint number: DESY 20-227, TTK-20-49.

\noindent\textit{Software:}
\textsc{emcee} \citep{emcee}, \textsc{NumPy} \citep{NumPy}, \textsc{Matplotlib} \citep{Matplotlib}, \textsc{Scikit-learn} \citep{scikit-learn}, \textsc{SciPy} \citep{SciPy}
%%%%%%%%%%%%%%%%%%%%%%%%%%%%%%%%%%%%%%%%%%%%%%%%%%
\section*{Data Availability}
The data underlying this article will be shared on reasonable request to the corresponding author.



\bibliographystyle{mnras}
\bibliography{references.bib}

%\begin{thebibliography}{99}
%\end{thebibliography}

%%%%%%%%%%%%%%%%%%%%%%%%%%%%%%%%%%%%%%%%%%%%%%%%%%

%%%%%%%%%%%%%%%%% APPENDICES %%%%%%%%%%%%%%%%%%%%%

\appendix

\section{Kernel overlap} \label{app:kernel_overlap}

In this appendix we discuss the computation of the kernel overlap $\Lambda_{ij}$, which arises from the integral of equation~\eqref{eq:drag_force2}.
To exploit symmetries, we express the integral in cylindrical coordinates,
\begin{multline} \label{eq:lambda_integral}
    \Lambda_{ij} = \, 2 \pi \int_{-\infty}^\infty \int_0^\infty W\left(\sqrt{z^2+r^2},h_i\right) \\ \times \, W\left(\sqrt{{\smash{(z-d)}\vphantom{z}}^2+r^2},h_j\right) \, r \, \mathrm{d}r \, \mathrm{d}z \, .
\end{multline}
In order to simplify the notation we have introduced the distance $d=|\Delta \mathbfit{x}|$ between the two particles. We have also integrated directly over the angle of the cylindrical coordinates, remaining only with two integrals over $r$ and $z$.
Using the assumption that the kernel function becomes zero beyond $h$, we can place tighter integration limits. For this purpose, we introduce $\gamma_i = \sqrt{h_i^2-z^2}$ and $\gamma_j = \sqrt{h_j^2 - (z-d)^2}$, such that
\begin{multline} \label{eq:lambda_integral_s0}
    \Lambda_{ij} = \, 2 \, \pi \int_{\max(-h_j,-h_j+d)}^{\min(h_i,h_j+d)} \int_0^{\min(\gamma_i,\gamma_j)} W\left(\sqrt{z^2+r^2},h_i\right) \\ \times \,  W\left(\sqrt{{\smash{(z-d)}\vphantom{z}}^2+r^2},h_j\right) \, r \, \mathrm{d}r \, \mathrm{d}z \, .
\end{multline}

For a given kernel function, $W(r,h)$ values for $\Lambda$ are tabulated in advance and then interpolated to obtain $\Lambda$ for given $h_i$, $h_j$ and $d$. Although $\Lambda$ depends on three variables, we need only a two-dimensional table for this purpose, as one variable can be interpreted as a scaling factor. To make this explicit, we introduce $h_\mathrm{min} = \min(h_i,h_j)$ and  $h_\mathrm{max} = \max(h_i,h_j)$ and scale all variables with $h_\mathrm{min}$, i.e.\ we introduce $d'= d/h_\mathrm{min}$ and $h' = h_\mathrm{max}/h_\mathrm{min}$. The scaled version of $\Lambda$ can then be written as
\begin{multline} \label{eq:lambda_scaled}
    \Lambda'_{ij}(d',h') = 2 \pi \int_{z'_\mathrm{min}}^{z'_\mathrm{max}}  \int_{r'_\mathrm{min}}^{r'_\mathrm{max}} W\left(\sqrt{{z'}^2+{r'}^2},1\right) \\ \times \, W\left(\sqrt{{\smash{(z'-d')}\vphantom{{r}}}^2+{r'}^2}, h' \right) r' \, \mathrm{d}r' \, \mathrm{d}z'
\end{multline}
with $r' = r/h_\mathrm{min}$ and $z' = z/h_\mathrm{min}$. The unscaled version of $\Lambda$ is then obtained from
\begin{equation} \label{eq:lambda_scaleing}
    \Lambda_{ij} = \frac{\Lambda'_{ij}}{h^3_\mathrm{min}} \, .
\end{equation}

\section{Implementation of rare self-interactions} \label{sec:rSIDM_implementation}
$N$-body simulations of DM with rare self-interactions employing an isotropic cross-section are well established.
There exists a variety of schemes, which differ in the way scatter probabilities are computed \citep[e.g.][]{Burkert_2000,Vogelsberger_2012,Rocha_2013}.
\cite{Rocha_2013} introduced a scheme where the scattering probability arises from the kernel overlap.
We follow this approach because we already compute overlaps for our fSIDM scheme.
By using the total cross-section $\sigma$ and the physical particle mass $m_\chi$ we can derive the scattering probability of a numerical particle pair.
Similar to the drag force we start from a microparticle travelling through a constant density $\rho$. The particle has the velocity $v$ and travels for the time $t$. The probability that it scatters with another particle is given by
\begin{equation} \label{eq:prob_scatter_rsidm}
	P_\mathrm{scatter} = \frac{\sigma}{m_\chi} \, \rho \, v \, t \,.
\end{equation}
Note, this is valid only for $P_\text{scatter} \ll 1$.
Now we consider two overlapping phase-space patches as represented by our numerical particles with densities $\rho_i$ and $\rho_j$. The expected number of scattering events is given as
\begin{equation}
	\langle N \rangle = \int \frac{\rho_i}{m_\chi} \, P_{\mathrm{scatter}} \, \mathrm{d}V \, .
\end{equation}
Here, $P_\mathrm{scatter}$ denotes the probability that a microparticle of $i$ scatters with one of $j$. We multiply by the microparticle mass and obtain the expected value for the mass per phase-space patch that scatters:
\begin{equation}
	\langle M \rangle = \frac{\sigma}{m_\chi} \, |\Delta \mathbf{v}_{ij}| \, \Delta t \int \rho_i \rho_j \mathrm{d}V \; ,
\end{equation}
where $\Delta \mathbfit{v}_{ij} = \mathbfit{v}_i - \mathbfit{v}_j$ is the relative velocity and $\Delta t$ is the simulation time-step.
The kernel overlap $\Lambda_{ij}$ is computed as described in Appendix~\ref{app:kernel_overlap}. We can then express the scattering probability of $i$ representing a mass of $m_i$ as
\begin{equation}
	P_i = \frac{\langle M \rangle}{m_i} = \frac{\sigma}{m_\chi} \, m_j \, |\Delta \mathbfit{v}_{ij}| \, \Delta t \, \Lambda_{ij} \, .
\end{equation}

For our implementation we use numerical particles that have the same mass $m$, such that $P_{ij} = P_i = P_j$.
The time-step $\Delta t$ is kept small enough such that the scattering probability is well below unity.
To determine whether two particles scatter during a given time-step we take a random number $x$ from the interval $[0,1]$ and let the particles scatter if $x \leq P_{ij}$.
The scattering process can be described as follows:
\begin{equation}
    \mathbfit{v}'_i = \mathbfit{v}_\mathrm{cms} + \frac{|\Delta \mathbfit{v}_{ij}|}{2} \, \mathbfit{e}
\end{equation}
and
\begin{equation}
    \mathbfit{v}'_j = \mathbfit{v}_\mathrm{cms} - \frac{|\Delta \mathbfit{v}_{ij}|}{2} \, \mathbfit{e} \, .
\end{equation}
Here, $\mathbfit{v}_\mathrm{cms} = (\mathbfit{v}_i + \mathbfit{v}_j)/2$, i.e.\ the centre-of-mass velocity.
The vector $\mathbfit{e}$ is a normalized vector that points into a random direction.
Here, we assume the cross-section to be isotropic, but anisotropic cross-sections can also be implemented \citep{Robertson_2017b}.
Our rSIDM implementation uses the same time-steps as for fSIDM (see Section~\ref{sec:adaptive_time_step}) and the same parallelization (see Section~\ref{sec:parallelisation}).

\section{Moli\`{e}re's theory} \label{ap:moliere}

In Section~\ref{sec:test_problem_deflection}, we use Moli\`{e}re's theory to predict the result of the angular deflection test problem.
Here, we give further details on how to derive the prediction.
The probability density distribution of the deflection angle, assuming scattering about small angles, was derived by \cite{Moliere_1948} and can be written as \footnote{Here, we only give the zeroth-order term, because the assumption of our method is that the underlying differential cross-section is extremely forward peaked, such that a given momentum transfer cross-section is achieved in the limit of an infinitely large cross-section for infinitesimally small-angled scattering events. In such a case, the so-called screening angle in Moliere theory is zero, which in turn implies that B \citep[see][]{Moliere_1948} is infinitely large, which means that the zeroth-order term is the only term that contributes to the distribution of scattering angles.}
\begin{equation} \label{eq:prob_total_deflection_angle2}
    f(\theta) = \frac{2 \, \theta}{\overline{\theta^2}} \, \exp\left( - \frac{\theta^2}{\overline{\theta^2}}\right) \,.
\end{equation}
To compute the distribution of the scattering angle $\theta$ one needs $\overline{\theta^2}$, which is given as
\begin{equation} \label{eq:moliere_omega2q_1}
    \overline{\theta^2} = 2 \pi \, n_0 \, l \, \int \frac{\mathrm{d} \sigma}{\mathrm{d} \Omega} \, \theta^3 \, \mathrm{d}\theta \, .
\end{equation}
Here, $n_0$ denotes the particle number density. It can be expressed as $n_0 = \rho/m_\chi$, where $\rho$ is the matter density and $m_\chi$ the physical particle mass.
The distance travelled by a particle through the target is given by $l$. Equation~\eqref{eq:moliere_omega2q_1} is not directly applicable for us as we only know the momentum transfer cross-section $\sigma_\mathrm{\tilde{T}}/m_\chi$.
But we can rewrite equation~\eqref{eq:moliere_omega2q_1} using the definition of the momentum transfer cross-section as given in \cite{Kahlhoefer_2014}.
\begin{align}
    \sigma_\mathrm{\tilde{T}} & = 4\pi \, \int_0^1 \frac{\mathrm{d}\sigma}{\mathrm{d}\Omega} \, (1-\cos \theta) \, \mathrm{d}\cos\theta \\
    &= -4\pi \int_{\pi/2}^0 \frac{\mathrm{d}\sigma}{\mathrm{d}\Omega} \, \sin \theta \, (1-\cos \theta) \, \mathrm{d}\theta \nonumber\\
    &\approx 2 \pi \int^{\pi/2}_0 \frac{\mathrm{d}\sigma}{\mathrm{d}\Omega} \, \theta^3 \, \mathrm{d}\theta . \label{eq:momentum_transfer_cross_section_approx}
\end{align}
In the final step we have assumed  that $\mathrm{d}\sigma/\mathrm{d}\Omega$ is strongly peaked at small angles, such that we can approximate $\sin\theta \, (1-\cos\theta) \approx \theta^3/2$. We therefore find
\begin{equation}
    \overline{\theta^2} \approx 2 \, n_0 \, l \, \sigma_\mathrm{\tilde{T}} = 2 \, \rho \, l \, \frac{\sigma_\mathrm{\tilde{T}}}{m_\chi} .
\end{equation}

\section{Stability of initial conditions}
\label{app:stability}

Here, we show that the NFW haloes used for our simulations in Sections~\ref{sec:halo_sim} and \ref{sec:merger_sim} are stable when evolved without DM self-interactions.

\begin{figure}
    \centering
    \includegraphics[width=\columnwidth]{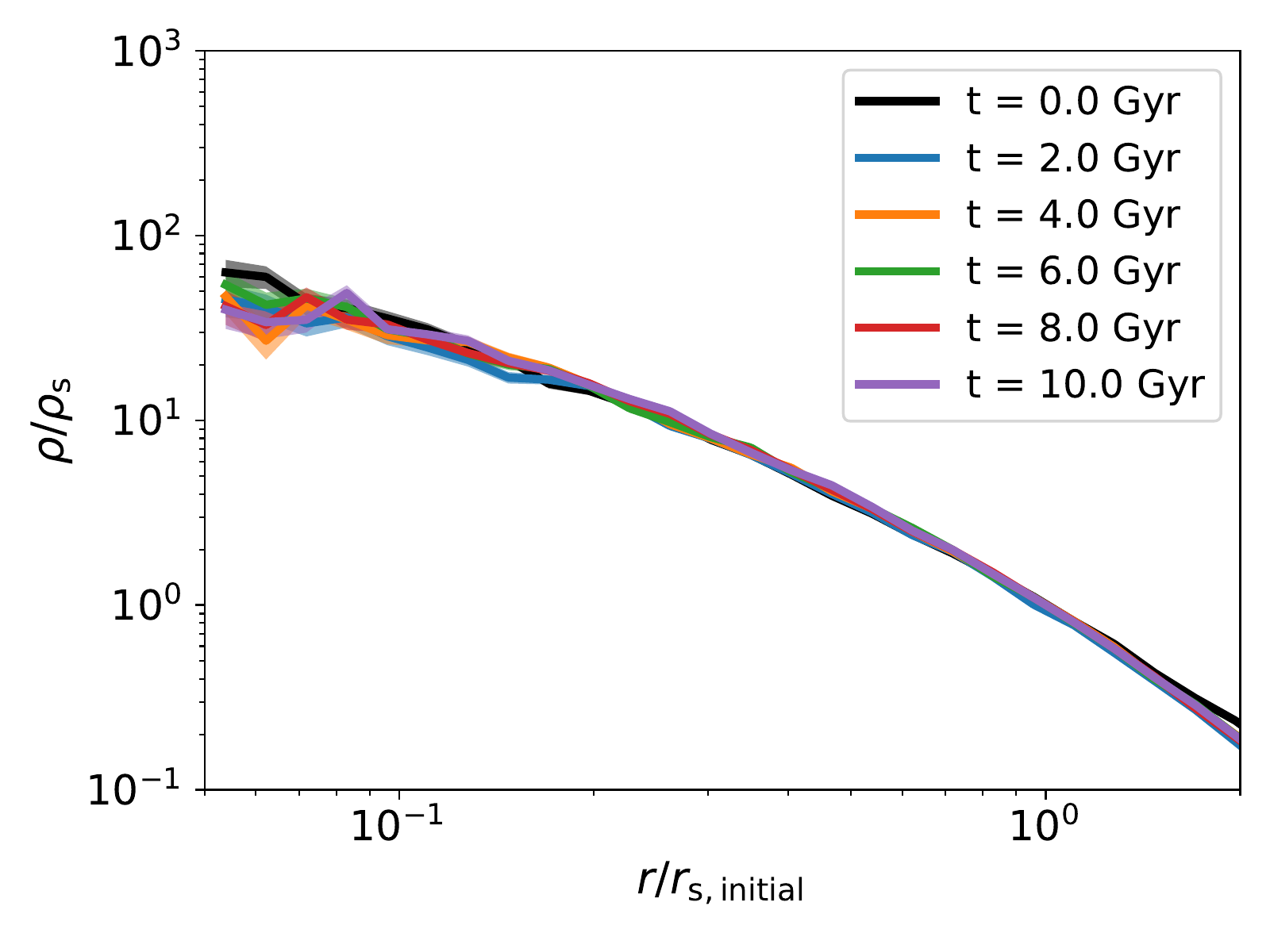}
    \includegraphics[width=\columnwidth]{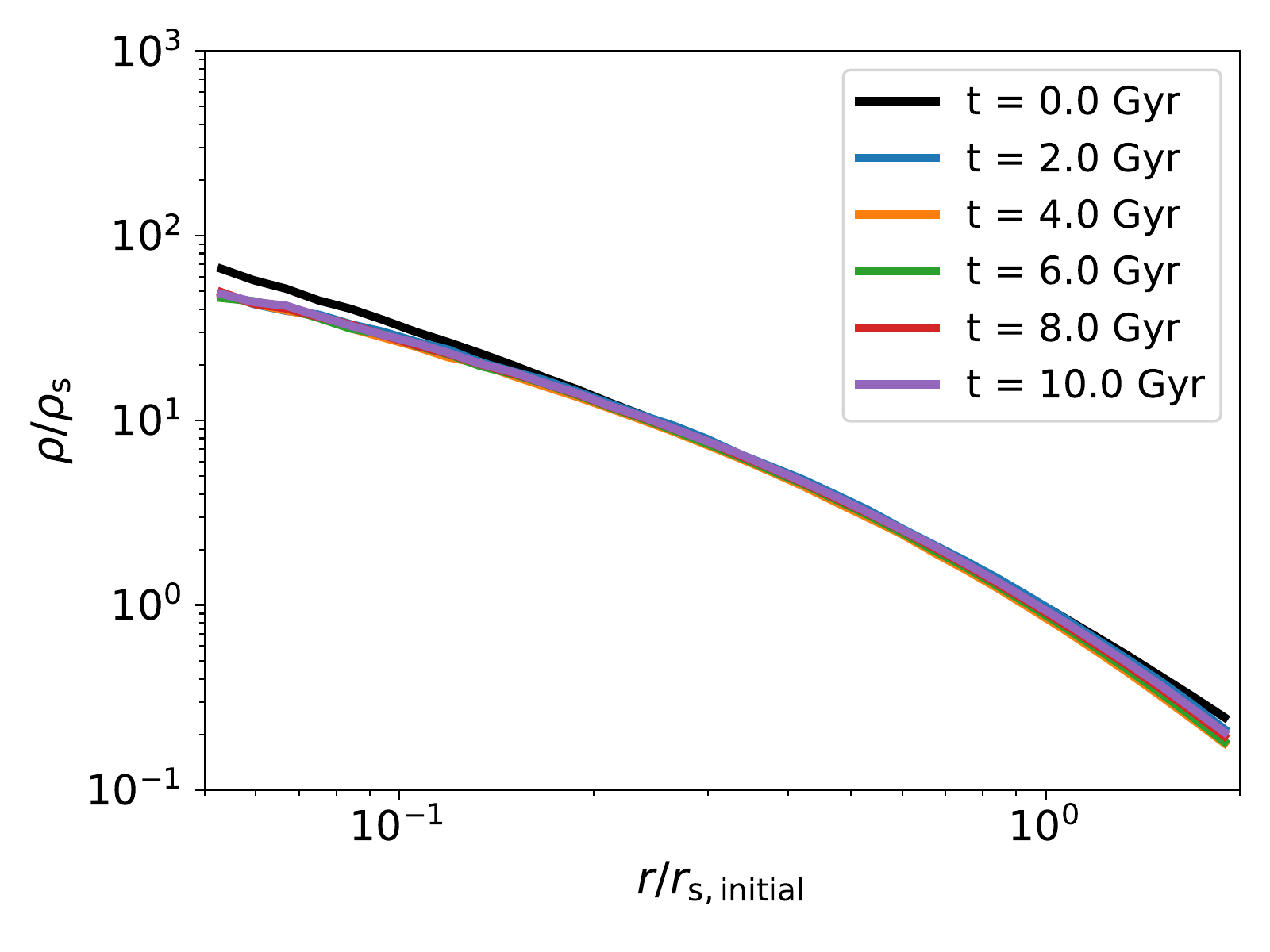}
    \caption{We show the evolution of an initial NFW halo as used for our test simulations in section~\ref{sec:halo_sim} (upper panel) and our merger simulations in section~\ref{sec:merger_sim} (lower panel). The haloes were simulated without DM self-interactions, i.e.\ consistent with CDM. Here, we display the density profile at several times.}
    \label{fig:halo_stab}
\end{figure}

For the simulations presented in Section~\ref{sec:halo_sim}, we used an initial NFW halo.
The halo has a virial mass of $M_\mathrm{vir} = 10^{15} \, \mathrm{M_\odot}$ and is resolved by $N=10^5$ particles.
In the upper panel of Fig.~\ref{fig:halo_stab}, we demonstrate the stability of these initial conditions.

In the lower panel of Fig.~\ref{fig:halo_stab}, we demonstrate that the haloes we use for our merger simulations (section~\ref{sec:merger_sim}) are stable when simulated without self-interactions.
One can only see minor changes of the density profile. The largest difference occurs in the centre of the halo. 

\section{Additional Merger Plots} \label{ap:merger_plots}
\begin{figure}
    \centering
    \includegraphics[width=\columnwidth]{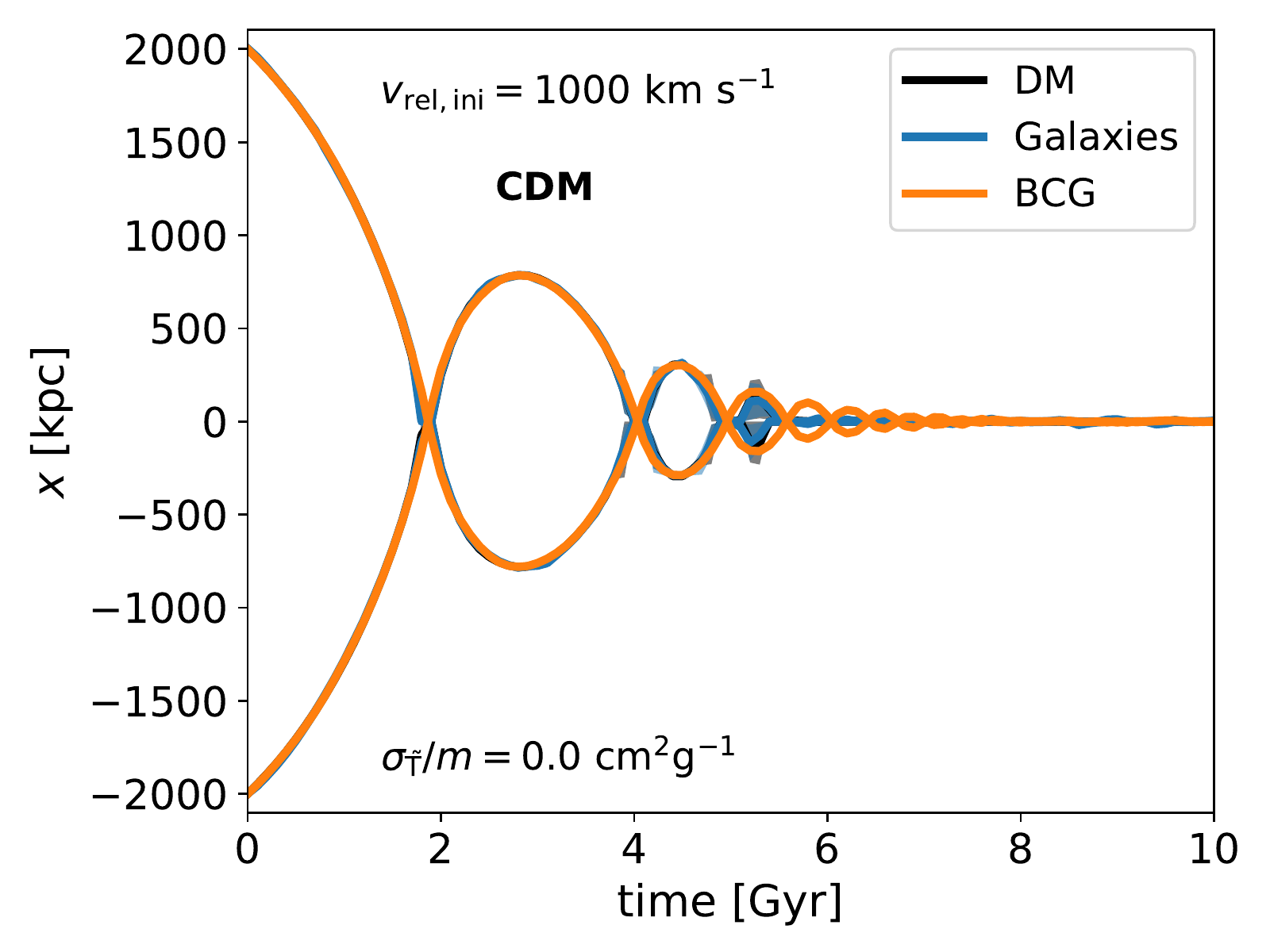}
    \caption{The same as in Fig.~\ref{fig:merger_byc_dist2} but for collisionless DM.}
    \label{fig:merger_byc_dist_cdm}
\end{figure}

\begin{figure}
    \centering
    \includegraphics[width=\columnwidth]{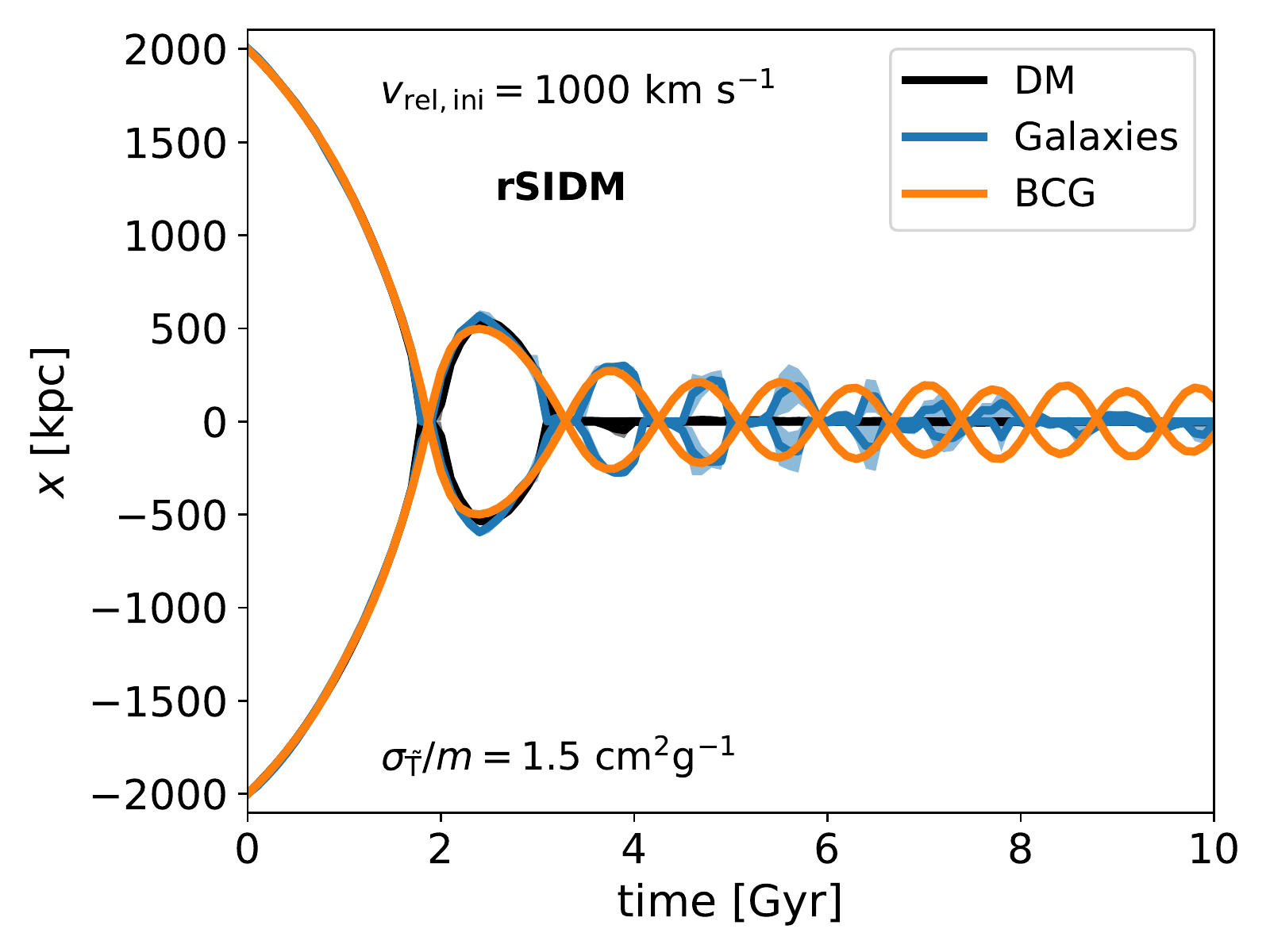}
    \caption{The same as in Fig.~\ref{fig:merger_byc_dist2} but for rare self-interactions.
    A self-interaction cross-section of $\sigma_\mathrm{\tilde{T}}/m = 1.5 \, \mathrm{cm}^2 \, \mathrm{g}^{-1}$ was employed.
    }
    \label{fig:merger_byc_dist2r}
\end{figure}

\begin{figure}
    \centering
    \includegraphics[width=\columnwidth]{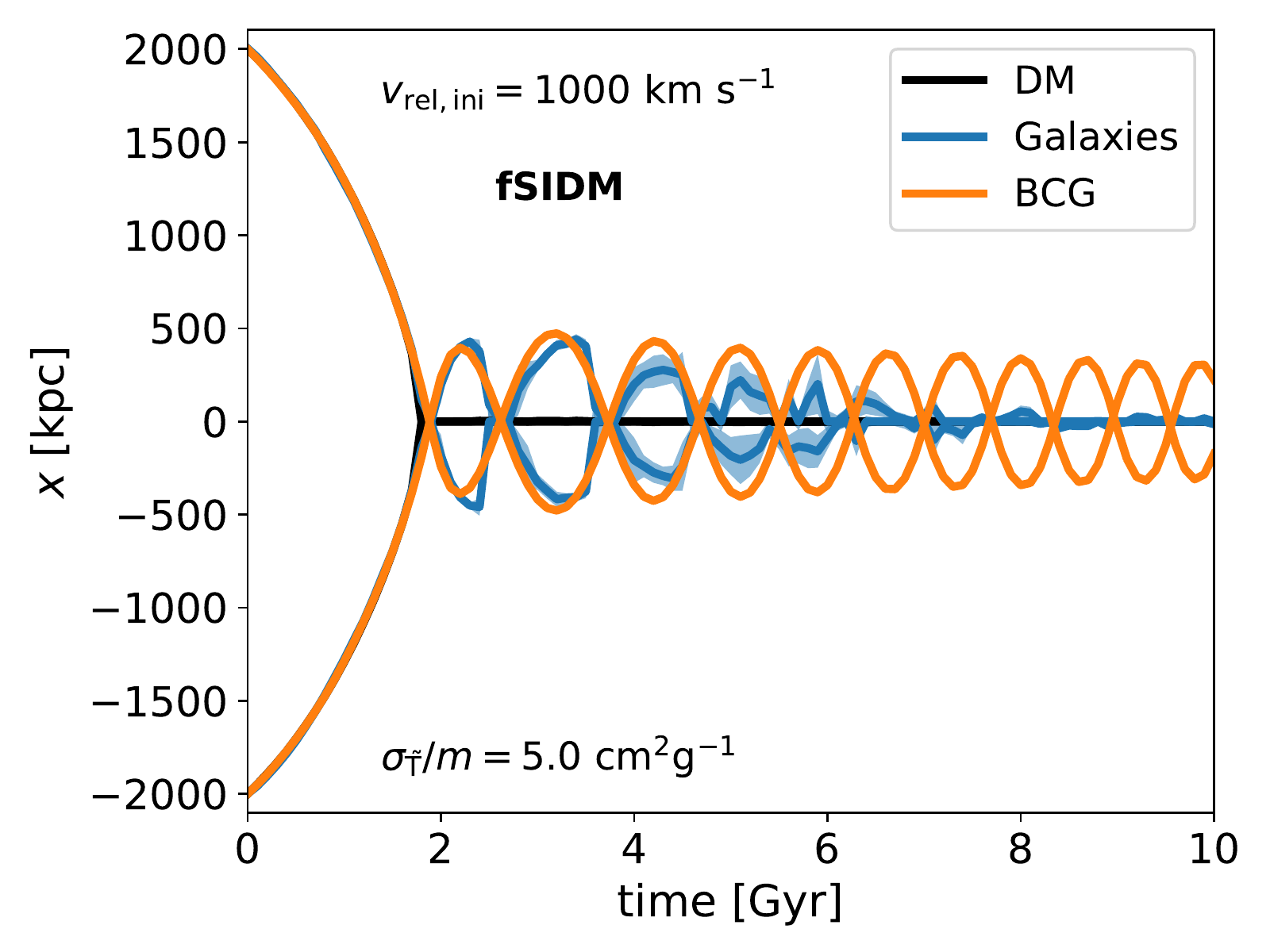}
    \includegraphics[width=\columnwidth]{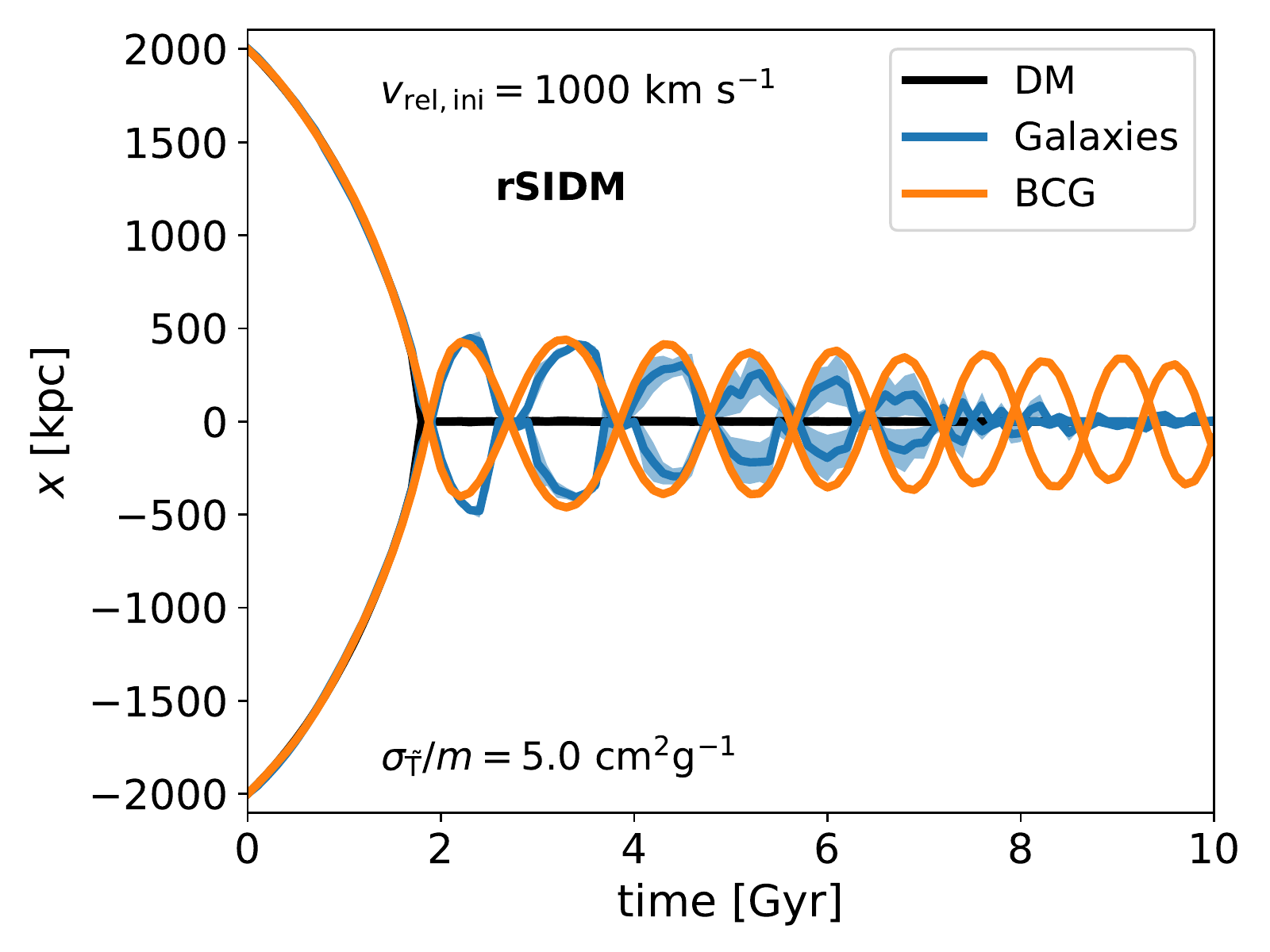}
    \caption{The same as in Fig.~\ref{fig:merger_byc_dist2}.
    The upper panel shows the evolution with frequent self-interactions and the lower panel displays the simulation with rare self-interactions.
    But for a cross-section of $\sigma_\mathrm{\tilde{T}}/m = 5 \, \mathrm{cm}^2 \, \mathrm{g}^{-1}$.
    Interestingly, the BCG peak distance at second apocentre is larger than at the first one. This is a consequence of the DM relaxation time. A flatter gravitational potential allows the BCG's to reach a larger distance at the second apocentre.}
    \label{fig:merger_byc_dist_10}
\end{figure}

Here, we show additional plots of our merger simulations, which are presented in section~\ref{sec:merger_sim}.
In particular, we show for peaks of all components the distance to barycentre as a function of time for cross-sections of $\sigma_\mathrm{\tilde{T}}/m = 0 \, \mathrm{cm}^2 \, \mathrm{g}^{-1}$ (Fig.~\ref{fig:merger_byc_dist_cdm}),
$\sigma_\mathrm{\tilde{T}}/m = 1.5 \, \mathrm{cm}^2 \, \mathrm{g}^{-1}$ (Fig.~\ref{fig:merger_byc_dist2r}),
and $\sigma_\mathrm{\tilde{T}}/m = 5 \, \mathrm{cm}^2 \, \mathrm{g}^{-1}$ (Fig.~\ref{fig:merger_byc_dist_10}).

\section{Amplification Process}
\label{ap:merger_amplification}

\begin{figure}
    \centering
    \includegraphics[width=\columnwidth]{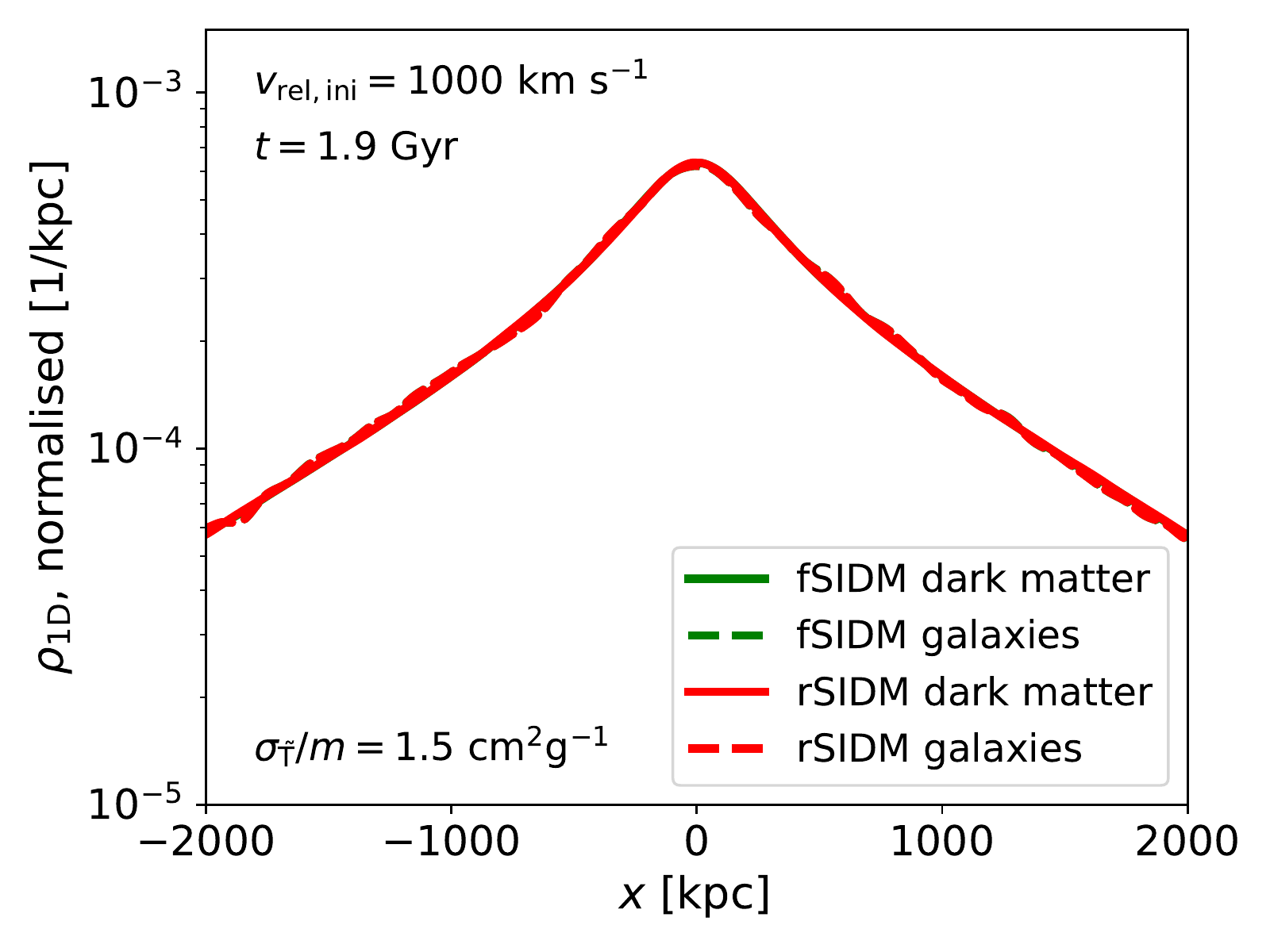}
    \includegraphics[width=\columnwidth]{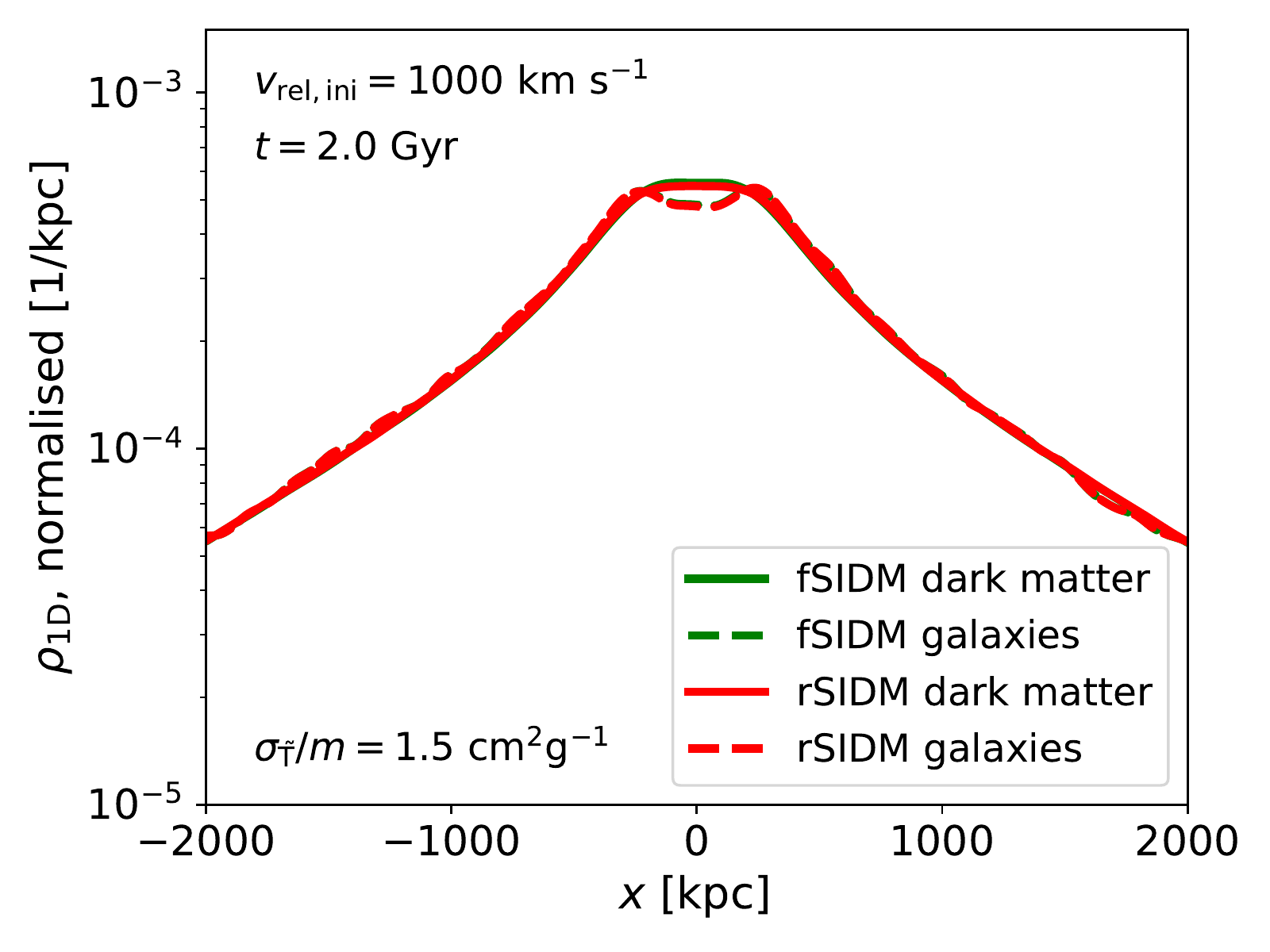}
    \includegraphics[width=\columnwidth]{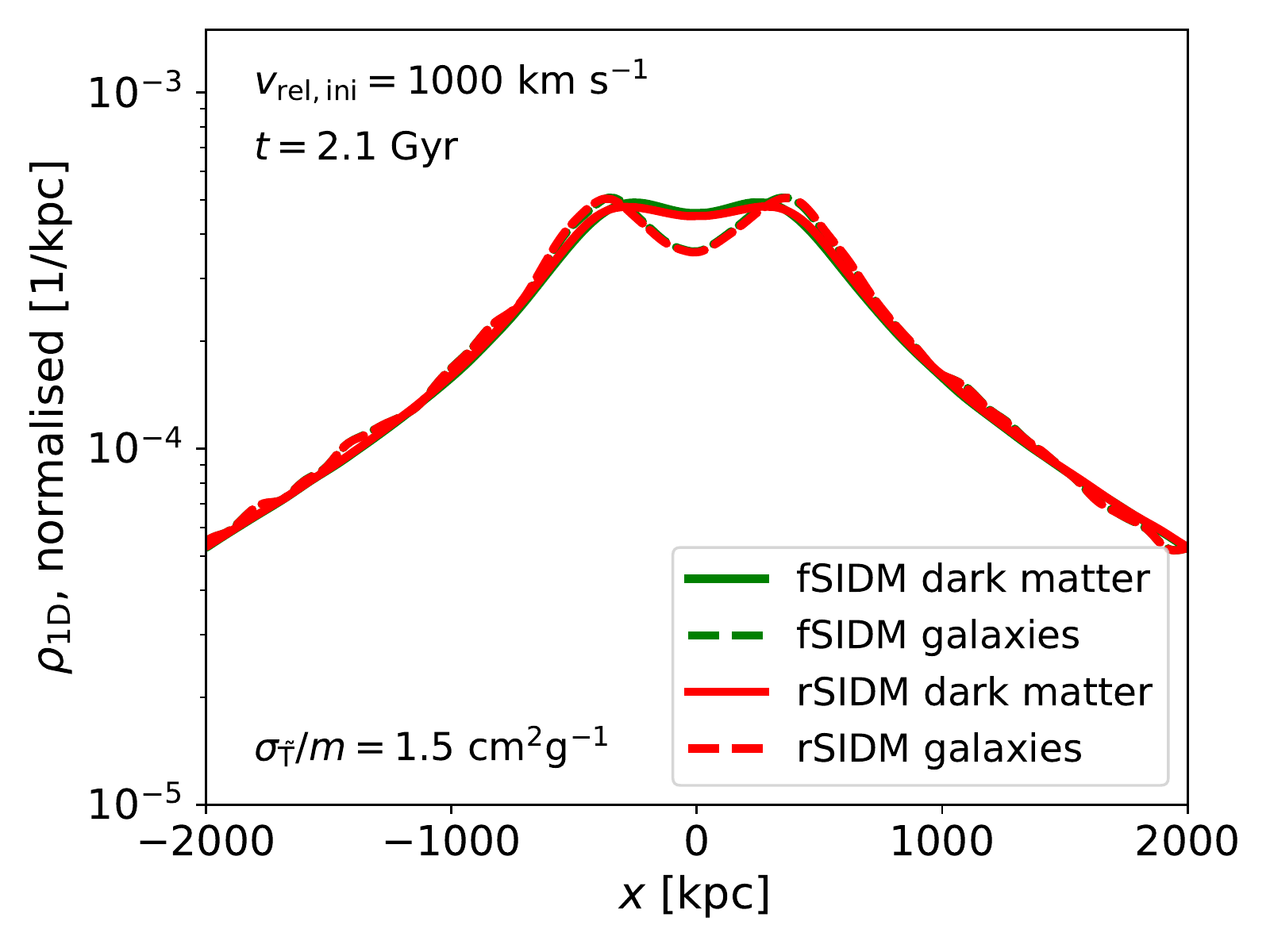}
    \caption{The normalized projected density along the merger axis is shown.
    We compare the density of the galactic and DM component for frequent and rare self-interactions.
    All three panels belong to the same cross-section ($\sigma_\mathrm{\tilde{T}}/m = 1.5 \, \mathrm{cm}^2 \, \mathrm{g}^{-1}$) and give the density for several times at pericentre passage and shortly afterwards.
    }
    \label{fig:merger_hist1d}
\end{figure}

\begin{figure}
    \centering
    \includegraphics[width=\columnwidth]{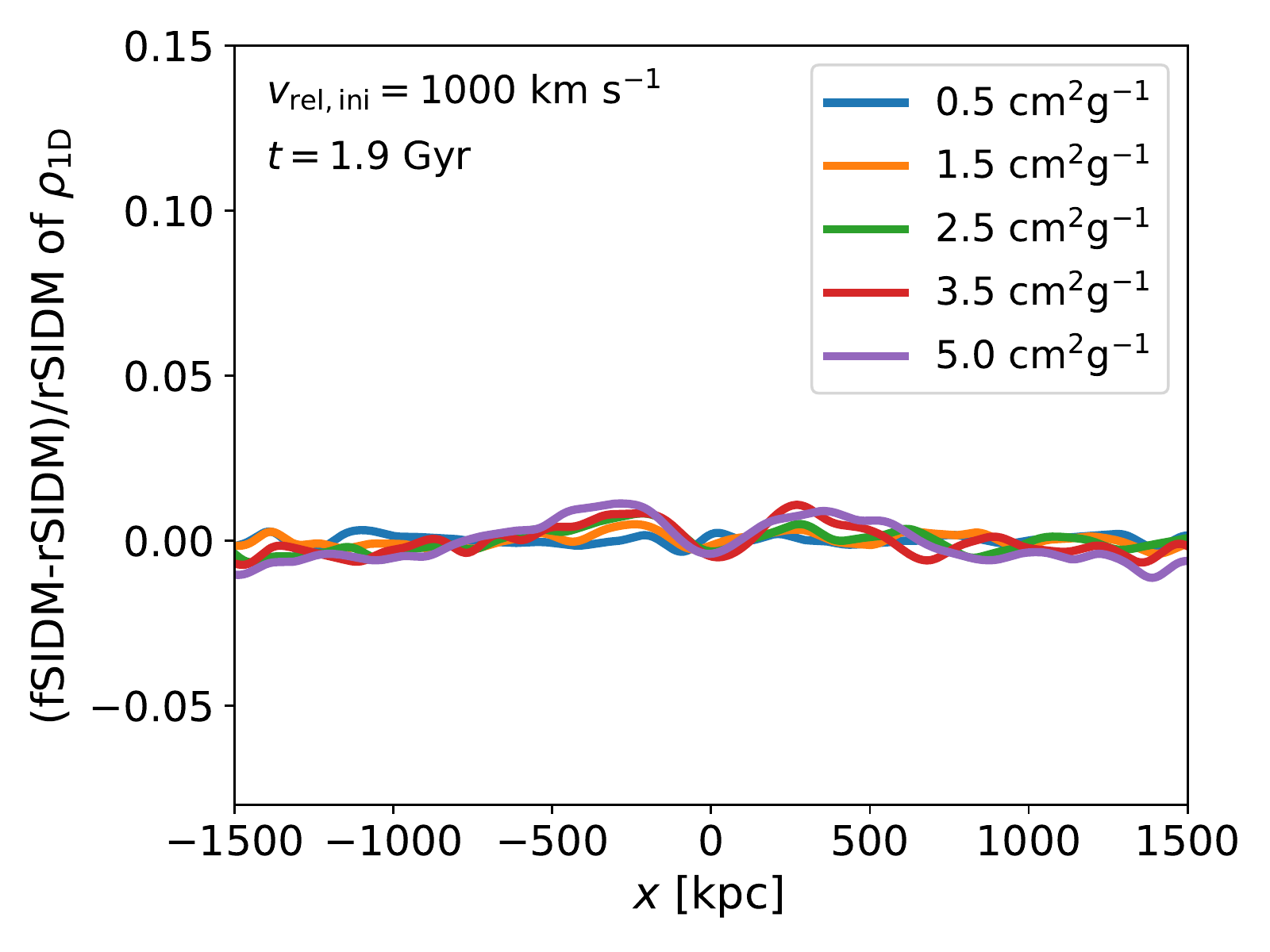}
    \includegraphics[width=\columnwidth]{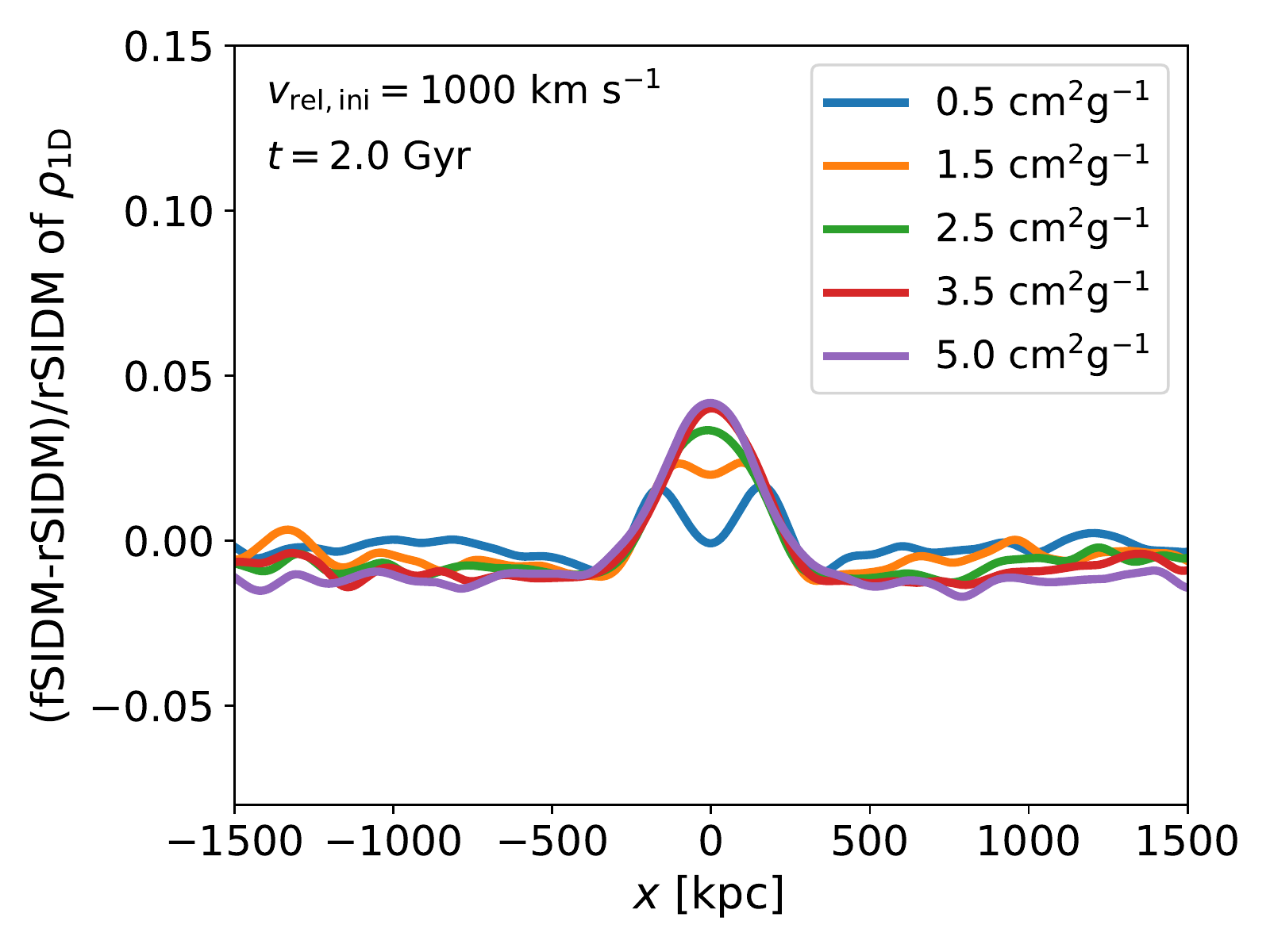}
    \includegraphics[width=\columnwidth]{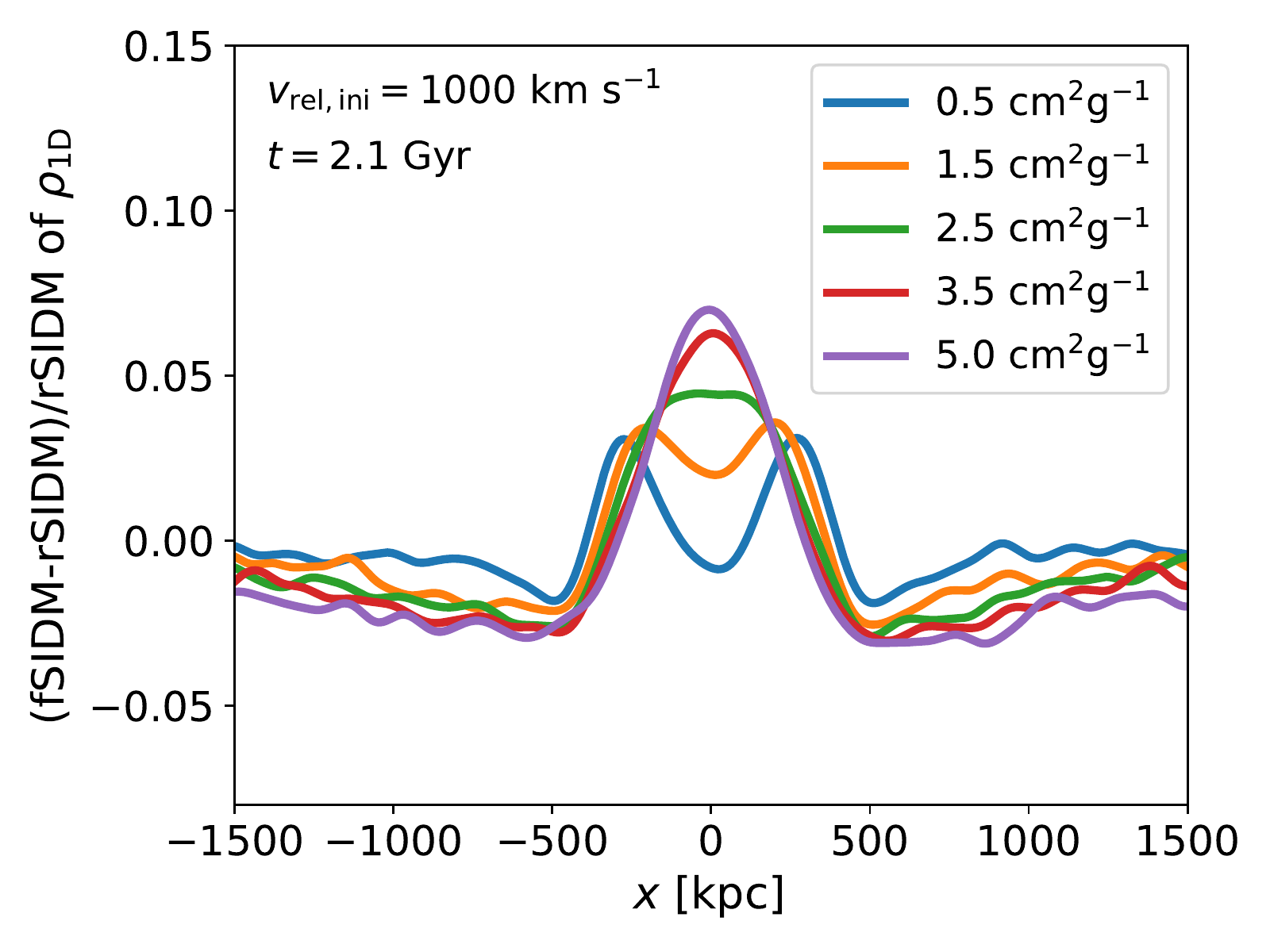}
    \caption{The relative projected density difference between fSIDM and rSIDM from Fig.~\ref{fig:merger_hist1d}, but for several cross-sections.
    A positive value implies that fSDIM is denser than rSIDM.}
    \label{fig:merger_dens_diff}
\end{figure}

As we mentioned in section~\ref{sec:merger_findings} the positions of BCGs and galaxy density peaks do not reflect the differences between rSIDM and fSIDM dark matter peaks one-to-one.
The deviation in BCGs and galaxies is larger than for the DM distribution when considering the maximum value of $\delta$ as shown in Fig.~\ref{fig:merger_peak_dev}.
In the following, we investigate this observation in more detail.

For this purpose, we compute the projected 1D density along the merger axis using a KDE with a 1D Gaussian smoothing kernel with a width of $50 \, \mathrm{kpc}$.
From this, we obtain the normalized density as shown in Fig.~\ref{fig:merger_hist1d}.
We also compute the difference between fSIDM and rSIDM, which is shown in Fig.~\ref{fig:merger_dens_diff}.

The key observation is that the central region close to the barycentre has a higher projected density for fSIDM than for rSIDM.
Although this could be a projection effect and does not necessarily imply that the actual density at the interaction point is larger for fSIDM, it clearly demonstrates that the distribution of DM, and hence the gravitational potential, differs for the two cases shortly after the collision.
This observation is readily understood in terms of the underlying differences between the two self-interaction schemes.
In fSIDM all DM particles are decelerated and deflected, i.e.\ some energy from the forward motion is redirected into the perpendicular direction.
In rSIDM, on the other hand, most DM particles are unaffected by self-interactions, while some particles scatter and experience a strong deflection. 

In rSIDM, the DM halo therefore travels further after pericentre passage than in fSIDM.
The deceleration of the DM component in fSIDM leads to a larger galaxy--DM offset.
The galaxies hence experience a stronger gravitational pull in fSIDM, which amplifies the differences in the galactic component between rSIDM and fSIDM.

%%%%%%%%%%%%%%%%%%%%%%%%%%%%%%%%%%%%%%%%%%%%%%%%%%

% Don't change these lines
\bsp	% typesetting comment
\label{lastpage}
\end{document}